\documentclass[twocolumn]{aastex7}
\usepackage[utf8]{inputenc}
\usepackage{amsmath}
\usepackage{amsfonts}
\usepackage{amsthm}
\usepackage{amssymb}
\usepackage{graphicx}
\usepackage{hyperref}
\usepackage{mathrsfs}%cursive font

\newcommand{\A}{\mathcal{A}}

\newcommand{\zhat}{\mathbf{\hat{z}}}

\newcommand{\ham}{\mathscr{H}}

\usepackage[normalem]{ulem}

\newif\ifcorr  
\corrtrue  % Comment this out (or switch to \purplefalse) to "turn off" purple text

\makeatletter
\DeclareRobustCommand{\volume}{\text{\volumedash}V}
\newcommand{\volumedash}{%
  \makebox[0pt][l]{%
    \ooalign{\hfil\hphantom{$\m@th V$}\hfil\cr\kern0.08em--\hfil\cr}%
  }%
}

\begin{document}

% \title{Mystery of Gaia’s Eccentric Wide Binaries Unraveled by Ménages à Trois in Star Clusters}
\title{A Million Three-body Binaries Caught by Gaia}

%\correspondingauthor{August Muench}
%\email{greg.schwarz@aas.org, gus.muench@aas.org}

\author[0000-0003-0136-8985]{Dany Atallah}
\affiliation{Department of Physics \& Astronomy, Northwestern University, Evanston, IL 60208, USA}
\affiliation{Center for Interdisciplinary Exploration \& Research in Astrophysics (CIERA), Northwestern University, Evanston, IL, 60201, USA}
\email{danyatallah@u.northwestern.edu}

\author[0000-0003-1992-1910]{Yonadav Barry Ginat}
\affiliation{Rudolf Peierls Centre for Theoretical Physics, University of Oxford, Parks Road, Oxford, OX1 3PU, United Kingdom}%
\affiliation{New College, Holywell Street, Oxford, OX1 3BN, United Kingdom}%
\email{yb.ginat@physics.ox.ac.uk}

\author[0000-0002-9660-9085]{Newlin C. Weatherford}
\affiliation{Observatories of the Carnegie Institution of Washington, 813 Santa Barbara Street, Pasadena, 91101, CA, USA}
\email{nweatherford@carnegiescience.edu}



\begin{abstract}
Gaia observations have revealed over a million stellar binary candidates within ${\sim}1\,{\rm kpc}$ of the Sun, predominantly characterized by orbital separations ${>}10^3\,{\rm AU}$ and eccentricities ${>}0.7$. The prevalence of such wide, eccentric binaries has proven challenging to explain through canonical binary formation channels. However, recent advances in our understanding of three-body binary formation (3BBF)---new binary assembly by the gravitational scattering of three unbound bodies (3UB)---have shown that 3BBF in star clusters can efficiently generate wide, highly eccentric binaries. We further explore this possibility by constructing a semi-analytic model of the Galactic binary population in the solar neighborhood, originating from 3BBF in star clusters and subsequently migrating to the solar neighborhood within a Hubble time. The model relies on 3BBF scattering experiments to determine how the 3BBF rate and resulting binary properties scale with local stellar density, velocity dispersion, and physically-motivated limits to 3UB encounters within a clusters' tidal field. The Galactic star cluster population is modeled by incorporating up-to-date prescriptions for the Galaxy's star formation history as well as the birth properties and internal evolution of its star clusters. Finally, we account for binary disruption induced by perturbations from stellar interactions before cluster dissolution and the subsequent changes and disruption of binary orbital elements induced by dynamical interactions in the Galactic field. Our model closely reproduces the total number of Gaia’s wide binaries and the separation and eccentricity distributions, suggesting that 3BBF may be an important formation channel for these enigmatic systems.

\end{abstract}


\section{Introduction} \label{sec:intro}
% \newlin{A general comment about the intro: you should try to better describe why it has been hard to explain wide, eccentric binaries and what the expectation is from canonical binary formation channels. Much of the heavy lifting here is done by simply citing chunky lists of papers, which leaves a lot of the burden on the reader to dig through these to contextualize the important problem/mystery you're trying to address. Some more exposition would be typical.}

The European Space Agency's \emph{Gaia} mission \citep{Gaia2016} has dramatically advanced our understanding of the Milky Way (MW)---\emph{inter alia}, it has discovered a multitude of wide, highly eccentric binaries, a combination of characteristics that has proven challenging to explain. Independent of origin, scattering encounters with other stars cannot drive binaries towards high eccentricities. Rather, such interactions tend to drive binaries toward thermal equilibrium, typically characterized by the \textit{thermal} eccentricity distribution $f(\epsilon) = 2\epsilon$ \citep{Jeans_1919,Ambartsumian1937,RoznerPerets2023,Makarov2025}. Since most stars and binaries are thought to have formed in dense stellar environments \citep{Zwart_2010}, where strong collisional dynamics are important, one might naively expect Gaia's wide binaries to be thermal. Instead, Gaia's observations suggest that wider binaries skew increasingly \textit{superthermal} (more eccentric) with increasing separations \citep{Tokovinin2020,ElBadry_2021,Hwangetal2022b,Hwang_2022}.

The superthermal nature of Gaia's wide binaries is even more puzzling when considering the impact of perturbations a binary would experience in the Galactic field. Though it can take longer than the age of the Universe to \emph{fully} thermalize a binary population in an environment as sparse as the Galactic field \citep{Gelleretal2019}, weak encounters with other bodies would still drive field binaries toward thermal eccentricities. Furthermore, \citet{ModakHamilton2023} proved that the secular effects from a general Galactic tide cannot transform a non-superthermal distribution into a superthermal one. Rather, these processes rapidly shift binaries towards thermal eccentricities on timescales that are expedited with increasing binary separation. Encounters in the disk serve to widen binaries, compounding the rate of binary thermalization.  Hence, Gaia's superthermal wide binaries must primarily arise at binary formation, with the original eccentricities skewed \emph{extremely} superthermal and semi-major axes (SMAs) smaller than those observed  \citep{Hamilton2022,ModakHamilton2023,Hamilton_2024}.

Metallicity may provide a further clue to the origin of Gaia's wide binaries. In particular, the two component members of each such binary typically exhibit near-identical metallicities, implying a co-natal formation channel \citep{Andrews_2018, Andrews_2019, Hawkins_2020}. In this vein, proposed formation channels for Gaia's wide binaries include cluster dissolution \citep{Kouwenhoven_2010, Moeckel_2010, Moeckel_2011}, the dynamical unfolding of triple systems \citep{Reipurth_2012}, random pairing of objects by thermodynamic fluctuations \citep{Penarrubia_2019} or primordially \citep{Marks_2012, Guszejnov_2023, Farias_2024}, dynamical interactions within tidal streams \citep{Penarrubia2021}, and primordial disk fragmentation \citep{Xu_2023}. Yet, these formation channels will either produce thermal (or sub-thermal) wide binaries at a consistent rate (e.g., random pairing) or, more rarely, binary populations that are only mildly superthermal (e.g., dynamical unfolding of triples and disk fragmentation). Whether these mechanisms may simultaneously produce extremely wide and highly eccentric binaries frequently enough to match the present-day populations in the solar neighborhood remains unclear due to a dearth of theoretical population studies accounting for Galactic scale wide-binary formation rates, Milky Way evolution, diffusion through the disk, and binary field evolution.

There is, however, a dynamical mechanism that readily and preferentially yields wide and highly eccentric binaries---three-body binary formation (3BBF). 3BBF is a triple encounter of initially unbound single stars, whose outcome is a bound binary and a more energetic single star \citep{Mansbach1970,AH76,Tutukov1978,Stodolkiewicz1986,Goodman_1993,ZwartMcMillan2000,HeggieHut2003,Pooleyetal2003,Ivanovaetal2005,Morscheretal2015,Weatherfordetal2023,Atallah_2024,Ginat_2024}. These new binaries---often termed `three-body binaries' (3BBs)---are expected to form frequently in stellar clusters. Binaries may also form in clusters by other means, such as primordial binary formation (e.g.~\citealt{Shuetal1987}) or tidal capture \citep{Fabianetal1975,PressTeukolsky1977}, but the fraction of eccentric wide binaries generated by these channels is significantly lower. This work investigates the capacity of 3BBF to generate the wide eccentric binaries observed by Gaia.  

Beginning with simulations of isolated scattering between three unbound bodies, we estimate the general 3BBF rate and resulting semi-major axis and eccentricity distributions. We then propagate these distributions in the context of a star cluster environment via a simplified astrophysical model, accounting for the chance of binary disruption until cluster dissolution, the dependence of these probabilities on cluster evolution, the MW's evolution, and the evolution of binaries deposited in the Galactic field. Our resulting formation rate, a function of nine parameters, predicts the number and orbital element distributions of binaries in the solar neighborhood that have formed via 3BBF in open (or larger) clusters. While 3BBF is automatically and self-consistently included in direct $N$-body simulations of star clusters (e.g.~\citealt{vanAlbada1968,Aarseth1969,BreenHeggie2012a,BreenHeggie2012b,Tanikawa2013,Wangetal2016,Parketal2017,Kumamotoetal2019,Arcaseddaetal2023}), these simulations are computationally prohibitive at this scale. In the spirit of versatility, we opt for a semi-analytic model.

Our methods and assumptions are outlined in \S \ref{sec:the Galactic 3BBF Rate: Fundamental Building Blocks}, leaving the full descriptions of each modeling component to the appendices. Specifically, we describe the treatment of 3BBF in star clusters in \S\ref{sec:Modeling Three-body Binary Formation in Star Clusters}, prescriptions for Galactic star formation history and radial diffusion in \S\ref{sec:Milky Way, Broad}, evolution of 3BBs in the Galactic field in \S\ref{sec:Field_evolution}, and the final piecing together of these components into a complete Galactic 3BBF rate (G3R) in  \S\ref{sec:G3R}. We then apply our model in \S\ref{sec:results} to predict the number of potentially observable 3BBs in the solar neighborhood and describe their semi-major axis and eccentricity distributions, among other properties. We finally discuss how our results compare to alternative binary formation channels, elaborate on key modeling uncertainties, possible future improvements, and lay out our conclusions in \S \ref{sec:summary}.

\begin{table*}
\centering
\caption{\textbf{G3R Independent Variables}: The independent, lowest-level integration variables used in the G3R in Equation~(\ref{eq:G3R_intro}).}
\label{tab:G3R Independent Variable}
  \centering
  \begin{tabular}{lll}
    \hline \hline
    Description & Symbol & First Mentioned \\
    \hline
    Radius of the interaction volume for 3BBF, normalized by local mean interparticle distance & $k\equiv R_1/r_{\rm sep}$ & \S~\ref{sec:The 3BBF Rate}\\

    Binary eccentricity & $\epsilon$  & \S~\ref{sec:The SMA/Eccentricity Distribution}\\
    
    Binary semi-major axis (SMA) & $a$  & \S~\ref{sec:The SMA/Eccentricity Distribution}\\

    Radial position in the host star cluster, normalized by the cluster's tidal radius & $\tilde{r} \equiv r/r_t$ & \S~\ref{sec:Plummer model}\\
    Time of 3BBF since the cluster's birth, normalized by the cluter's total lifetime & $\tilde{\tau}\equiv \tau / \tau_{\rm cl}$ & \S~\ref{sec:cluster evolution} \\
    Initial cluster mass & $M_{\rm cl0}$ & \S~\ref{sec:cluster evolution} \\
    Galactocentric radius of the cluster's orbit and intial radius of the escaping 3BB & $R_G$ & \S~\ref{sec:cluster evolution}/\ref{sec:Milky Way Evolution}\\
    Galactocentric radius of the binary's orbit at time $t+\tau$ & $R_G'$ & \S~\ref{sec:Milky Way Evolution}  \\
    Time between Milky Way birth and cluster birth & $t$ & \S~\ref{sec:Milky Way Evolution}\\
  
    \hline
  \end{tabular}
\end{table*}

\section{The Galactic 3BBF Rate: Fundamental Building Blocks}
\label{sec:the Galactic 3BBF Rate: Fundamental Building Blocks}
Accurately predicting the contribution of 3BBF in dynamically active star clusters to the population of binaries in the Galactic field requires the careful assembly of a comprehensive set of distribution functions. These distribution functions describe the properties and evolution of 3BBs and the star clusters that produce them across MW history. When combined, the final equation may be used to estimate the G3R, the total rate of 3BBF within the Galaxy. This section broadly outlines the philosophy of our calculation and the separable components of the G3R, described further in \S\ref{sec:G3R}.

By combining each distribution, the G3R is expressible as a differential rate with respect to nine key integration variables:
\begin{equation} \label{eq:G3R_intro}
\begin{aligned}
    \mathcal{G}_{\rm MW}&=\frac{dN_{\rm bin}}{d\Omega} =  \frac{\volume_*}{\volume_{\rm R_{G,s}}}[\tilde{\Gamma}_F f_F P_{\rm no \, enc} f_{\rm CIMF} \\
    &\quad \times \left( N_{\rm ha} f_{\rm ha} SFH_{\rm ha} D_{\rm ha} + N_{\rm la} f_{\rm la} SFH_{\rm la} D_{\rm la}\right)]\\
    d\Omega&= r_{\rm t}\left(M_{\rm cl0}, \tilde{\tau}, R_G \right) \, \tau_{\rm cl}(M_{\rm cl0})\\
        & \quad   \times\, d\tilde{r} \, d\tilde{\tau} \, dk \, \, dM_{\rm cl0}\,dR_G  \, dR_G' \, dt \, \, d\epsilon \, da  
\end{aligned}
\end{equation}
where $N_{\rm bin}$ is the total number of 3BBs and $\volume_*/\volume_{\rm R_{G,s}} = \delta l^2/(3 R_{\rm G,s} h) \approx 4.2\times 10^{-4}$ is the fraction of a $\pm 100\,\rm{pc}$ annulus with thickness $h \approx 1\rm kpc$ at solar Galactocentric distance $R_{G,s}=8 \,\rm{kpc}$ (i.e., an annulus from Galactocentric radii $7.9$--$8.1\,\rm{kpc}$) that lies within a $\delta l=100\,\rm{pc}$ sphere centered on the Sun. This coefficient is necessary to reduce our model dataset to the solar neighborhood, in accord with Gaia's observational limits. The rough detection limit for Gaia binaries is ${\sim}1\,\rm{kpc}$, but we limit our comparison region because Gaia binary detections become highly incomplete beyond $200\,\rm{pc}$ \citep{ElBadry_2021}. For ease of reference, Table~\ref{tab:G3R Independent Variable} lists each of the nine independent integration variables in this equation while Table~\ref{tab:G3R Building Blocks} lists the separable distribution functions listed alongside the sections where they are discussed in detail. 

Binary field evolution, whether by secular evolution or scattering encounters in the disk, is applied to 3BBs that escape their natal clusters and migrate to the present-day solar neighborhood. The two field evolution prescriptions are ``phase mixing'' (\textsc{PM}, \S~\ref{sec:phasemixanddiskdisruption}) and ``cumulative scatter'' (\textsc{CS}, \S~\ref{sec:cumulative scatter}) and are denoted by the operators $\mathcal{F_{\rm pm}}$ and $\mathcal{F_{\rm cs}}$, respectively. To obtain a prediction for the distribution of Gaia binaries, we numerically integrate (via the Monte Carlo method) the expression
\begin{equation}
    N_{\rm bin} = \int \mathcal{F}\left(\mathcal{G}_{\rm MW}\right) d\Omega.
\end{equation}
Computationally, the operators are incorporated after Monte Carlo sampling of $\mathcal{G}_{MW}$. The operators simply act to (non-conservatively) transform the original distribution, of which Monte Carlo sampling serves to uncover.

Listed below are brief descriptions and motivations surrounding the building blocks:

\begin{table*}
\centering
\caption{\textbf{G3R Building Blocks}: The separable distribution functions forming the basis of the G3R in Equation~(\ref{eq:G3R_intro}).}
\label{tab:G3R Building Blocks}
  \centering
  \begin{tabular}{llll}
    \hline \hline
    Name & Symbol & Required Variables & Location \\
    \hline
    Volumetric 3BBF Rate & $\tilde{\Gamma}_{\rm F}$ & $\tilde{r}, \tilde{\tau}, M_{\rm cl0}, R_G, k$ & \S \ref{sec:The 3BBF Rate} \\
    Binary semi-major axis \& eccentricity distribution & $f_{\rm F}$ & $\tilde{r}, \tilde{\tau}, M_{\rm cl0}, R_G, \epsilon, a,  k$ & \S \ref{sec:The SMA/Eccentricity Distribution} \\
    % Binary escape probability & $P_{\rm esc}$ & $\tilde{r}, \tilde{\tau}, M_{\rm cl0}, R_G, a$ & \S \ref{sec:escape_disruption_cluster} \\
    Binary in-cluster survival probability & $P_{\rm no \, enc}$& $\tilde{r}, \tilde{\tau}, M_{\rm cl0}, R_G, a$ & \S \ref{sec:escape_disruption_cluster} \\
    Initial stellar distribution in disk & $f_{\rm d}$ & $R_G$ & \S \ref{sec:Milky Way Evolution} \\
    Star formation history of disk & $SFH_{\rm d}$ & $R_G, t$ & \S \ref{sec:Milky Way Evolution} \\
    Stellar diffusion distribution through disk & $D_{\rm d}$ & $\tilde{\tau}, M_{\rm cl0}, R_G, R_G', t$ & \S \ref{sec:Milky Way Evolution} \\
    Cluster initial mass function & $f_{\rm CIMF}$ & $M_{\rm cl0}, R_G, t$ & \S \ref{sec:cimf} \\
    Binary survival probability in disk (only if phase mixing) & $P_{\rm d, MW}$ & $\tilde{\tau}, M_{\rm cl0}, R_G, R_G', t, a$ & \S \ref{sec:disk disruption} \\
    \hline
  \end{tabular}
\end{table*}
 \begin{enumerate}

\item  $\tilde{\Gamma}_F$: The foundational element of the calculation is the semi-analytic, volumetric 3BBF rate for the case of three equal-mass bodies, $\tilde{\Gamma}_F$, as derived in \citet{Atallah_2024} and numerically evaluated with higher accuracy in \S~\ref{sec:The 3BBF Rate}. Every other component of the G3R will serve to ``project'' the volumetric 3BBF rate across time and Galactic location and to account for the destruction of 3BBs given where and when they are born in a cluster and the cluster's location in the Galactic disk. 

\item $f_{\rm F}$: The two-dimensional distribution of binary eccentricity and semi-major axis (SMA) for 3BBF between bodies with equal masses and velocities, $f_{\rm F}$, is described in \S~\ref{sec:The SMA/Eccentricity Distribution}. This distribution is independent of the volumetric rate, normalizing to unity over the entire parameter space of possible binary orbital parameters.

% \item $P_{\rm esc}$: The fraction of 3BBs, formed with a velocity greater than their host cluster's local escape speed, assuming an isolated cluster, $P_{\rm esc}$, is calculated from the 3BB velocity distribution in \S~\ref{sec:escape_disruption_cluster}. The (Maxwellian) 3BB velocity dispersion is the geometric sum of the cluster's local velocity dispersion and the velocity kick imparted to the binary from the potential energy released during 3BBF. Naturally, the kick is small, especially for (locally) soft/wide binaries, but the escape rate of hard/tight binaries would be substantially underestimated if the kick were neglected.

\item $P_{\rm no \, enc}$: Only a small fraction of wide 3BBs will avoid a perturbing encounter or total ionization (i.e., disruption via encounters with other stars) by the time the cluster dissolves. The binary's probability of surviving without undergoing an encounter, $P_{\rm no \, enc}$, is calculated in \S~\ref{sec:escape_disruption_cluster} and is conservatively assumed to be given by the geometric encounter rate as we are only interested in the fraction of binaries unperturbed by any encounters. We do not consider the alternative---where the binary receives a strong kick upon formation, allowing it to escape from the cluster---for this event relies on the exponentially unlikely event that three unbound bodies with velocities at or in excess of the escape velocity (the tail of the distribution function) may meet in a small volume and form soft binaries with a center-of-mass velocity in excess of the local escape velocity. Additionally, the equal-mass 3UB simulations used to generate the semi-analytic fits of \S~\ref{sec:The SMA/Eccentricity Distribution} generate wide binaries with CoM velocities approximately a factor of $\sqrt{2}$ slower than the initial Maxwellian by which the initial velocities were drawn.  
\item $(f_d \times SFH_d \times D_d)$: Moving to considerations external to the host star cluster, our MW model describes the radial distribution of star clusters in the MW, when they are born, and the rate at which 3BBs leave their clusters to migrate through the disk to the solar neighborhood. Detailed in \S~\ref{sec:Milky Way Evolution}, our MW model is a combination of the semi-analytic frameworks designed by \citet{Frankel_2019,Frankel_2020}, while also separating the thin and thick disks according to \citet{Wagg_2022}. 

The first of the three separable disk components is $f_d$---a one-dimensional distribution encoding the initial Galactocentric radius, $R_G$, of stars/clusters born in the MW. The star formation history, $SFH_d$, is the distribution of stars born in the disk as a function of time $t$ (and $R_G$ for the thin disk, though the function is still only normalized in time). The diffusion function, $D_d$, characterizes the resulting Galactocentric radius, $R_G'$, of the binary after radial migration over time $\tau$ since the cluster's birth. The subscript `d' is replaced with `ha'/`la' when tuned for the MW thick/thin disk, respectively.

These three building blocks are tuned with observationally determined scaling constants, separately characterizing the thin and thick disks. Each function individually (and all collectively) integrates to unity: $f_d$ over all initial Galactocentric radii, $SFH_d$ over time and up to the present age of the MW (assumed to be 12 Gyrs), and $D_d$ over all Galactocentric radii at time of 3BBF. It is here that we multiply our linearly separable disk distributions by the number of stars in each disk. Thus, integrating over the entire disk parameter space using the disk building blocks yields the total number of stars in each present-day disk.

\item $f_{\rm CIMF}$: Our MW model distributes all stars born at time $t$ and location $R_G$ into clusters of initial mass $M_{\rm cl0}$ (also born at time $t$) based on the cluster initial mass function (CIMF), $f_{\rm CIMF}$. The CIMF incorporates a hybrid of the \citet{Just_2023} CIMF and a \textit{Schechter} cutoff \citep{Schechter_1976} that is dependent on the local star formation rate at $R_G$ in the disk---implicitly found using the MW model described above. It integrates to unity between $50\,M_\odot$ and the Schechter upper mass cutoff, $M_c$; see \S~\ref{sec:cimf} for details.

\item $P_{\rm d,MW}$: Finally, two field binary evolution prescriptions, ``phase mixing'' \citep{HamiltonRafokiv2019II,Hamilton2022} and ``cumulative scatter'' \citep{Hamilton_2024}, are described and applied to model binaries as post-processing in \S~\ref{sec:Field_evolution}. The final building block, $P_{\rm d,MW}$, calculates the fraction of 3BBs that are not disrupted while migrating through the disk, but \emph{is only relevant when phase mixing} (\textsc{PM}). \textsc{PM} evolves the binary eccentricity secularly, imparting a small effect on binary eccentricity as it travels the MW disk. On the other hand, ``cumulative scatter'' (\textsc{CS}) simulates the evolution of a binary due to the cumulative perturbations of stellar encounters in the disk, self-consistently accounting for binary orbital evolution and disruption. \textsc{CS} is a far stronger effect, imparting a dramatic shift in the eccentricity and SMA of 3BBs, strongly perturbing the mock distributions into alignment with observations. The derivation of the disruption rate and discussion of the post-processing prescriptions are found in \S~\ref{sec:Field_evolution}.

\end{enumerate}

%%%%%%%%%%%%%%%%%%%%%%%%%%%%%%%%%%%%%%%%%%

%%%%%%%%%%%%%%%%%%%%%%%%%%%%%%%%%%%%%%%

%%%%%%%%%%%%%%%%%%%%%%%%%%%%%%%%%%%%%%%%%

\begin{figure*}
 \gridline{
        \fig{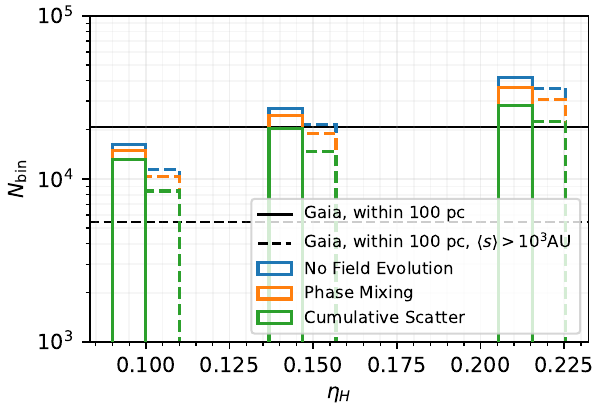}{0.48\textwidth}{(i)}
        \fig{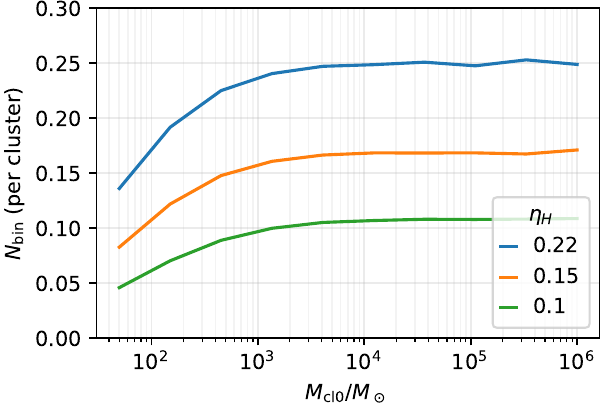}{0.48\textwidth}{(ii)}}
    \caption{\textit{Left}: Total number of three-body binaries, $N_{\rm bin}$, we predict are presently within $100\,\rm pc$ of the Sun. We do not account for any observational incompleteness in our models, aside from limiting our comparison radius. The black solid line shows the total number of Gaia binaries observed within $100\,\rm pc$ \citep{ElBadry_2021}, only including the Gaia binary candidates assessed to have a ${<}10\%$ probability of being a chance alignment of two unbound stars. The dashed lines are the totals having removed binaries with average separations ${<}10^3 \rm AU$. Our models exclude 3BBF under too strong a tidal field within a cluster using the ``tidal isolation'' parameter, $\eta_H$; it sets the maximum allowed tidal acceleration applied by the cluster to the three scattering bodies (Equation~\ref{eq:R1_RH}; see also \S~\ref{sec:The 3BBF Rate}). Color denotes the field evolution prescription, phase mixing (\textsc{PM}, \S~\ref{sec:phasemixanddiskdisruption}) and cumulative scatter (\textsc{CS}, \S~\ref{sec:cumulative scatter}). The \textsc{CS} prescription destroys more binaries because the destruction and scattering rate of binaries tends to increase following each subsequent encounter and is the unconditionally dominant effect when compared to \textsc{PM}. \textit{Right}: The average number, $N_{\rm bin}$, of three-body-binaries (3BBs) that survive without undergoing an encounter in a cluster of initial mass $M_{\rm cl0}$ until cluster dissolution; this is largely independent of Galactocentric radius. Note that the right panel does not consider binary disruption in the MW disk or the subset of binaries that migrate to the solar neighborhood. A very small fraction of the initial cluster population forms 3BBs that survive till dissolution without undergoing an encounter, with ${\lesssim}25\%$ of clusters emitting a single unperturbed 3BB to the field.} 
    \label{fig:Nbin.pdf}
\end{figure*}

\begin{figure*}
 \fig{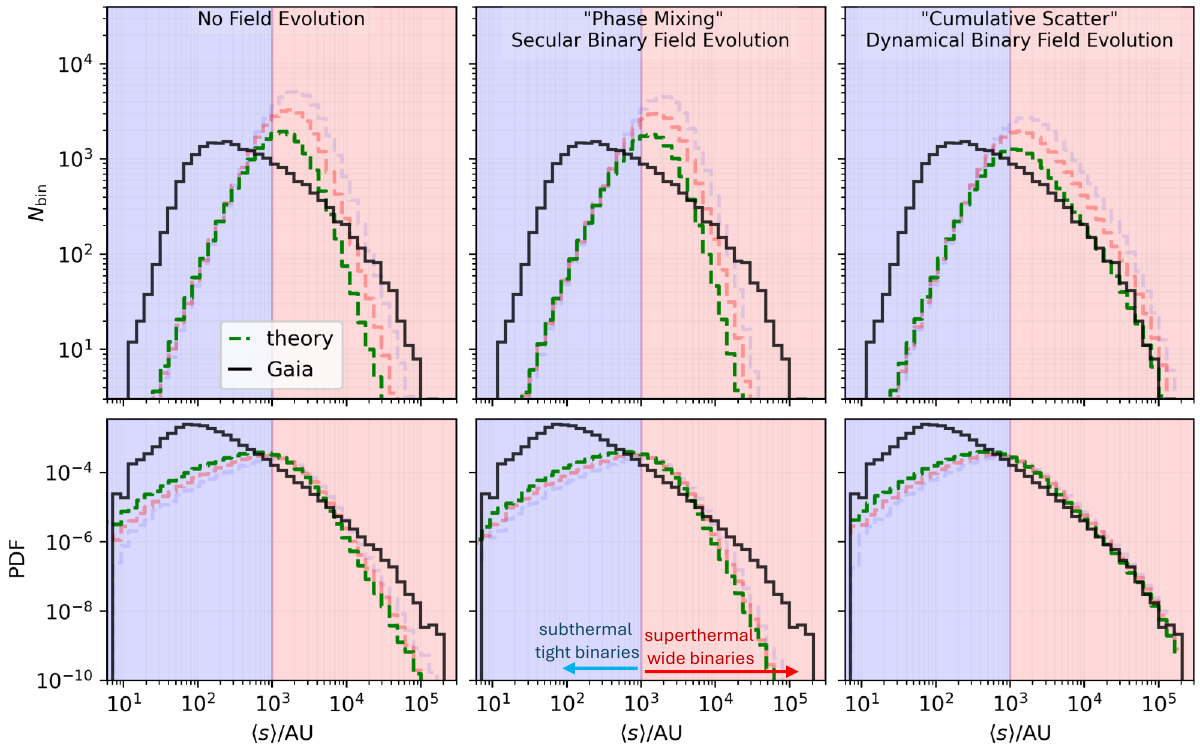}{\textwidth}{}
    \caption{The predicted number, $N_{\rm bin}$ (\textit{top}), of 3BB's within $100 \, \rm pc$ of the Sun and their probability density function (\textit{bottom}) as a function of $\langle s\rangle$, the projected average binary separation (Equation~\ref{eq:sep}).\footnote{Each colored curve is a model realization with a different ``tidal isolation'' parameter, $\eta_{H}$ (see \S \ref{sec:Gaia sma and ecc}). Our most conservative value, $\eta_H=0.1$, is the dashed purple line, denoting a local tidal acceleration $0.1\%$ of the average relative acceleration between bodies during three-body scattering. Larger $\eta_H$ yields more binaries, but risks breaking the model assumption of isolated three-body scattering.} The Gaia binary separations within 100 pc of the Sun are (black curve) taken from \citet{ElBadry_2021} using their ${<}0.1$ binary chance-alignment criterion. The \emph{left} panel is the final distribution when applying ``phase mixing'' (\textsc{PM}, \ref{sec:phasemixanddiskdisruption}) while the \emph{right} panel uses the ``cumulative scatter'' prescription  (\textsc{CS}, \ref{sec:cumulative scatter}). The \textsc{CS} prescription represents the most physically accurate model. It is highly improbable that a wide binary can secularly evolve in isolation over the course of the MW's entire history, as is assumed in the \textit{PM} model. The overall shape of the separation distribution closely matches the observed distribution when $\langle s\rangle>10^3 \rm AU$. The total number (i.e., from summing over $N_{\rm bin}$) of 3BBs is reported in Figure~\ref{fig:Nbin.pdf}.} 
    \label{fig:Gaia_sma_all.pdf}
\end{figure*}

\begin{figure*}
    \fig{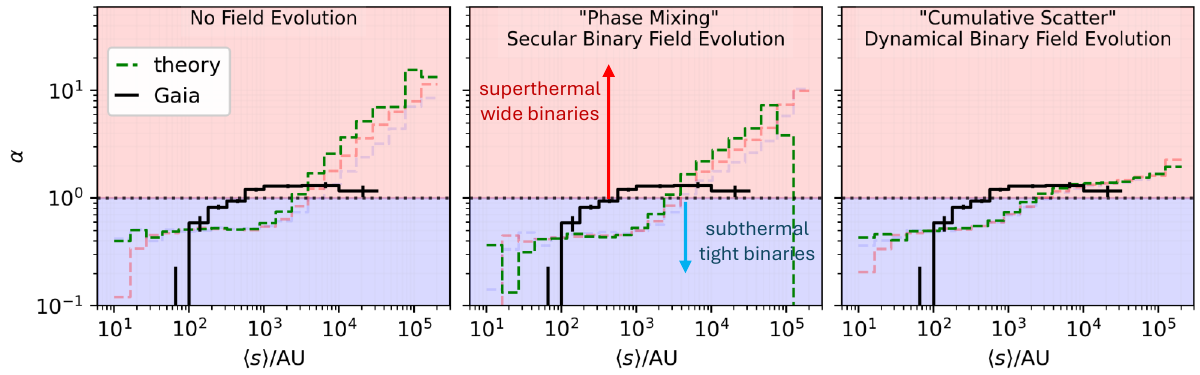}{\textwidth}{}
    \caption{The eccentricity parameter $\alpha \equiv \mathrm{d}\ln f/\mathrm{d}\ln \epsilon$ (Equation~\ref{eq:ecc_alpha_fit}), as a function of the projected average binary separation, $\langle s\rangle$. Note that $\alpha=1$ corresponds to the thermal distribution, while $\alpha>1$ and  $\alpha <1$ represent superthermal and subthermal distributions, respectively.\footnote{Each colored curve denotes the value of $\eta_{\rm H}$ selected to integrate the G3R (Equation~\ref{eq:G3R}); $\eta_{\rm H}$ is the maximum size of the spatial three-body interaction region (defined in \S \ref{sec:Gaia sma and ecc}) in units of the triple's Hill radius. Our preferred model, $\eta_H=0.10$ (green), corresponds to a maximum scattering region $10\%$ the local Hill Radius within a cluster. Our most lax condition, $\eta_H=0.22$ (light blue), represents a tidal acceleration $1\%$ of the average relative acceleration between bodies.} The solid black line is the set of $\alpha$ values found by \citet{Hwang_2022}. In the \emph{left} panel the colored dashed lines are the ``phase-mixed'' distributions (\S~\ref{sec:phasemix}). The \emph{right} panel features the final eccentricity distributions following the ``cumulative scatter'' prescription (\S~\ref{sec:cumulative scatter}). Phase mixing imparts only a slight ``thermalization,'' with the resultant wide binary distributions far more eccentric than what is observed. Rather, by considering dynamical encounters in the field, cumulative scattering imparts a dramatic thermalization, resulting in eccentricities in greater observational agreement with \citet{Hwang_2022}. Despite the simplicity of our toy model and the uncertainties surrounding Gaia-incompleteness, there is good qualitative agreement: wide binaries tend superthermal while tight binaries tend subthermal.}
    \label{fig:gaia_ecc_all.pdf}
\end{figure*}

\section{Mock Three-body Binary Population versus Gaia Wide Binaries}\label{sec:results}

The entirety of our results stem from evaluating and interpreting the G3R (\S~\ref{sec:G3R}, Equation~\ref{eq:G3R}), following the application of field evolution post-processing (\S~\ref{sec:Field_evolution}), and comparing to Gaia observations of binaries as compiled by \citet{ElBadry_2021} and \citet{Hwang_2022}. Despite the several idealizations necessary to implement our semi-analytic model (e.g., equal-mass stars, universal Plummer clusters, circular and unchanging cluster orbits in the MW, etc.), our results share remarkable agreement with Gaia's wide binary population in SMA and eccentricity distributions, and the total number within the solar neighborhood.

Figure~\ref{fig:Nbin.pdf} displays the total number of solar neighborhood 3BBs we predict across all our models, alongside the wide binary $\left (\langle s\rangle>10^3 \rm AU\right)$ tally. The ``tidal isolation'' parameter, $\eta_H$, is a nuisance parameter quantifying how isolated a three-body encounter is from tidal perturbations applied by its host cluster.  As later discussed in \S~\ref{sec:The 3BBF Rate},  the interaction volume is defined to be 
\begin{equation}\label{eq:R1_RH}
  R_1 = \eta_H R_H,  
\end{equation}
where $R_H$ is the local Hill Radius. Shrinking $\eta_H$ further isolates three-body encounters, with $\eta_H= \{0.22, 0.1\}$ excluding 3BBF if the maximum tidal acceleration applied to the three bodies by the cluster during the encounter is ${>}\{1\%, 0.1\%\}$ of the three bodies' maximum initial relative acceleration. 

Though we may increase certainty in the physical ``isolation'' of a 3BBF event by shrinking $\eta_{\rm H}$, what constitutes ``isolated enough'' is unconstrained in the $N$-body literature. Several $\eta_H$ models are generated to show the effect of asserting ever more conservative tidal isolation during 3BBF. A clear trend presents itself in Figure~\ref{fig:Gaia_sma_all.pdf}: smaller $\eta_H$ dramatically constrains the 3BB creation probability, especially for the widest binaries. This is a direct result of the 3BBF rate scaling by ${\sim}\eta_{\rm H}^3$ (\S~\ref{sec:The 3BBF Rate}). Depending on $\eta_H$, the number of 3BBs our model predicts to currently be within $100\,\rm{pc}$ ($1\,\rm{kpc}$) of the Sun varies from ${\sim}10^4$--$10^5$ (${\sim}10^6$--$10^7$).

The choice of field evolution prescription heavily impacts the number and orbital element distributions of field binaries, with \textsc{CS} applying the most significant perturbations to individual binaries and the binary population as a whole. That said, both, the \textsc{PM} and \textsc{CS} binary field evolution prescriptions help bring the initial, highly superthermal 3BB eccentricities in line with the more modestly superthermal observations. This is consistent with the conclusions of \citet{Hamilton2022,Hamilton_2024}, who postulated that the physical mechanism(s) responsible for creating the observed Gaia wide binaries would need to have a dramatic superthermal bias because of the thermalizing effect stellar encounters and the Galactic tide impart following binary formation.

These 3BBs are the rare few that make a clean getaway---undergoing no encounters---from their natal cluster after formation. From Figure~\ref{fig:Nbin.pdf}, the expected number of 3BBs escaping to the Galactic disk over a cluster's life asymptotes to $0.2$--$0.5$. The 3BBF creation and survival rates asymptote because the overwhelming majority of binaries that survive till dissolution are necessarily assembled late in the cluster's life, when there are fewer than ${\sim}100$ bodies. Although three‑body binary formation (3BBF) events are intrinsically rare and the resulting binaries most often fail to survive within any single cluster, the sheer abundance of low‑mass clusters compensates for this scarcity. The demographics of these binaries are discussed further in \S\S~\ref{sec:Gaia sma and ecc} and \ref{sec:3bbf formation characteristics}, where the former section explores 3BBF's capacity to reproduce the observable Gaia superthermal wide binary population, and the latter examines the predicted solar-neighborhood binary population characteristics.

\subsection{Semi-major Axis and Eccentricity Distributions}\label{sec:Gaia sma and ecc}

To further compare the modeled and Gaia-observed binary populations, we express their respective SMA ($a$) and eccentricity ($\epsilon$)\footnote{As a reminder to the reader, we employ $\epsilon$ as the eccentricity, not the eccentricity squared as is usually the case in the literature.} distributions in Figures~\ref{fig:Gaia_sma_all.pdf} and \ref{fig:gaia_ecc_all.pdf} in terms of the average projected separation \citep{Hwang_2022}:
\begin{equation}\label{eq:sep}
\langle s \rangle=\frac{\pi}{4} a \left(1 - \frac{\epsilon^2}{2}\right).
\end{equation}
Figure~\ref{fig:Gaia_sma_all.pdf} provides the most visually compelling agreement, especially when we conservatively limit our maximum allowable 3BBF encounter volume to ${\sim}10\%$ of the local cluster Hill Radius ($\eta_H=0.1$). According to our model, the suppression in the wide binary population at higher $\langle s \rangle$ is directly attributable to binary disruption by disk encounters (Equation~\ref{eq:disk disruption probability}). The 3BBF ${\sim}a^{2.5}$ proportionality for hard binaries, as expressed in \citet{Heggie_1975, Goodman_1993} and numerically reproduced in \S~\ref{sec:The SMA/Eccentricity Distribution}, mildly correlates with the observed Gaia curve for separations ${<}10^3\,\rm{AU}$. However, there is substantial evidence that the tight‐binary regime is highly incomplete \citep{Elbadry_2018}, so any apparent agreement is most likely coincidental, implying that 3BBF can only play a supporting role in directly assembling tight binaries.

The orbital perturbations by disk encounters on 3BBs is modeled with greater physical realism in the \textsc{CS} prescription. It does not merely consider the absolute destruction or conservation of binaries evolving under constant, isolated, secular evolution by the Galactic tides (as in \textsc{PM}), but enables stochastic binary evolution in SMA and eccentricity. A gentler fall-off in the separations of the wide 3BB population results, with most binaries with final SMA ${>}10^4 \rm AU$ born tighter, with initial SMA between $10^3-10^{3.5} \rm AU$ (discussed further in \S~\ref{sec:3bbf formation characteristics}). These tighter binaries end up populating the widest subset of binaries through \textsc{CS} evolution, mitigating losses due to the rapid destruction of binaries with initial SMA $>10^4 \rm AU$.

Turning to binary eccentricities,  Figure~\ref{fig:gaia_ecc_all.pdf} draws a set of fits for the eccentricity parameter, $\alpha$, with our two different field evolution prescriptions (\S~\ref{sec:Field_evolution}) and overlaid by the Gaia values calculated in \citet{Hwang_2022}. The $\alpha$ parameter is a convenient means to quantify how the eccentricity distribution compares to the thermal distribution. By fitting the expression
\begin{equation}\label{eq:ecc_alpha_fit}
f(\epsilon) = (\alpha+1) \epsilon^\alpha,
\end{equation}
the $\alpha$ distribution is subthermal (less eccentric than thermal) if $\alpha<1$, thermal when $\alpha=1$, and superthermal if $\alpha>1$. It should be immediately apparent that the predicted and observed $\alpha$ values are not an exact match (variances within order unity), however they are in good qualitative agreement. Echoing the observational findings of \citet{Hwang_2022}, our models show that binaries with $\langle s\rangle<10^3 \rm{AU}$ tend subthermal while binaries with $\langle s\rangle>10^3 \rm{AU}$ tend superthermal. The subthermal behavior is expected for binaries that form near the peak of the two-dimensional SMA--eccentricity distribution (where $a\approx R_1$, the radius of the three-body interaction volume; see \S~\ref{sec:The SMA/Eccentricity Distribution}).

The degree to which wide binaries exhibit a superthermal distribution varies dramatically between the \textsc{PM} and \textsc{CS} field evolution prescriptions (left panel and right panels of Figure~\ref{fig:gaia_ecc_all.pdf}, respectively). Because the \textsc{PM} prescription only provides a cursory thermalizing effect, the eccentricity distributions of the widest binaries remain unreasonably superthermal. To the contrary, \textsc{CS} produces an $\alpha$ distribution in closer agreement with observations for wide binaries. As $\eta_H$ shrinks, both field evolution methods produce more eccentric binaries, though this is effect is small relative to how dramatically binary SMA and formation rates are affected by $\eta_H$.

\begin{figure*}
    \centering
    % Adjust width (e.g., 0.8\columnwidth or 0.9\textwidth) as needed.
    \includegraphics[width=\textwidth]{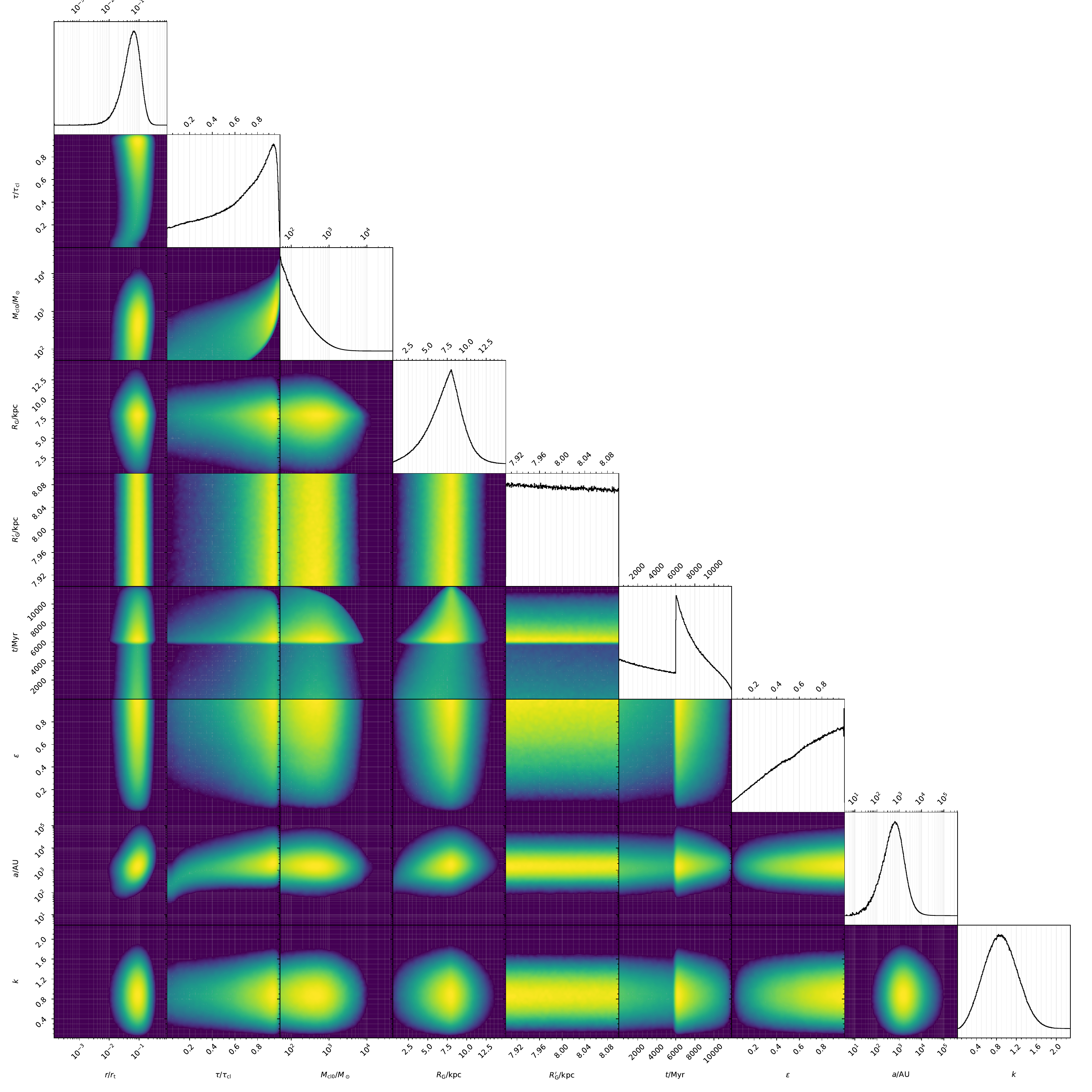}
    \caption{Grid of the resultant 9D, Galactic 3BBF Rate (G3R, Equation~\ref{eq:G3R}) distribution, as obtained by Monte Carlo integration with \textsc{emcee} \citep{emcee}, and projected as a set of 2D probability density functions for each dimensional combination. See Table~\ref{tab:G3R Independent Variable} for a description of the independent variables.  The sharp discontinuity in the $t$ dimension begins at the onset of star formation in the MW thin disk. This model was generated by fixing the maximum encounter volume to $10\%$ of the local cluster Hill Radius ($\eta_{\rm H}=0.1$, \S~\ref{sec:The 3BBF Rate}) and evolving field  binaries with the ``cumulative scatter'' (\textsc{CS}) prescription (\S~\ref{sec:cumulative scatter}). The grid was generated using the Python \textsc{corner} module \citep{corner}.}
    \label{fig:corner.pdf}
\end{figure*}

\begin{figure*}
    % First plot
    \gridline{
        \fig{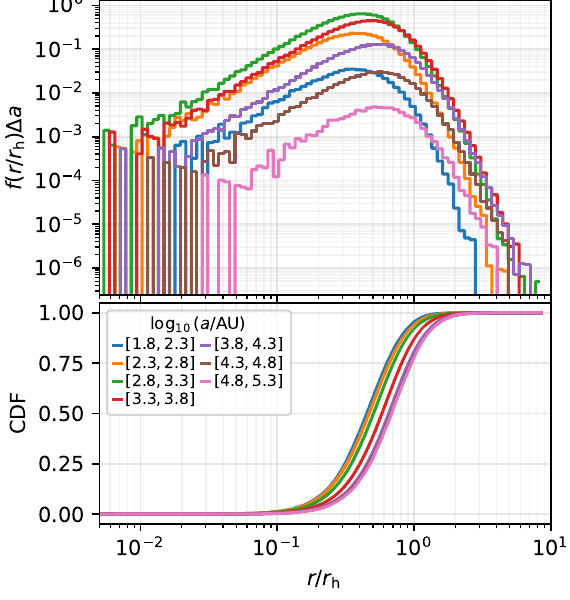}{0.321\textwidth}{(a)}
        \fig{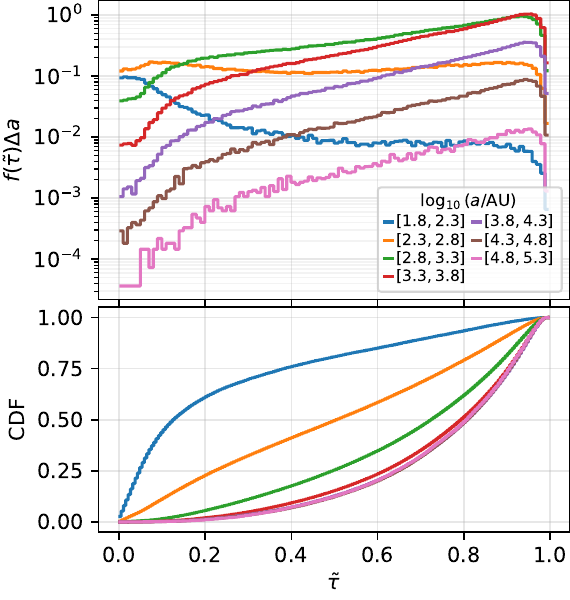}{0.321\textwidth}{(b)}
        \fig{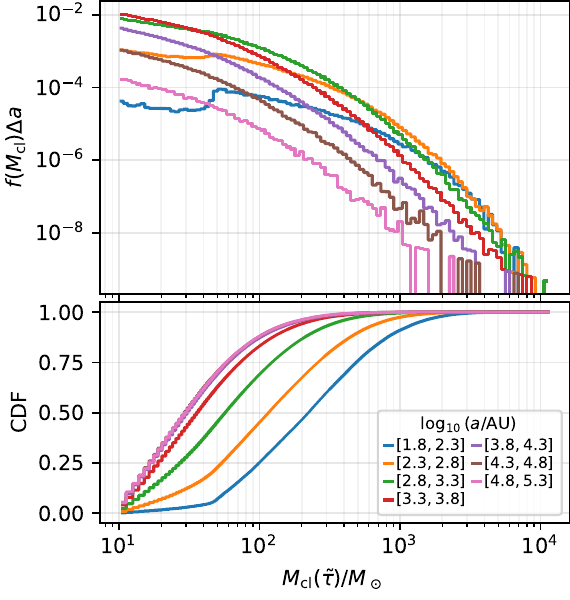}{0.321\textwidth}{(c)}
        }
    \gridline{
        \fig{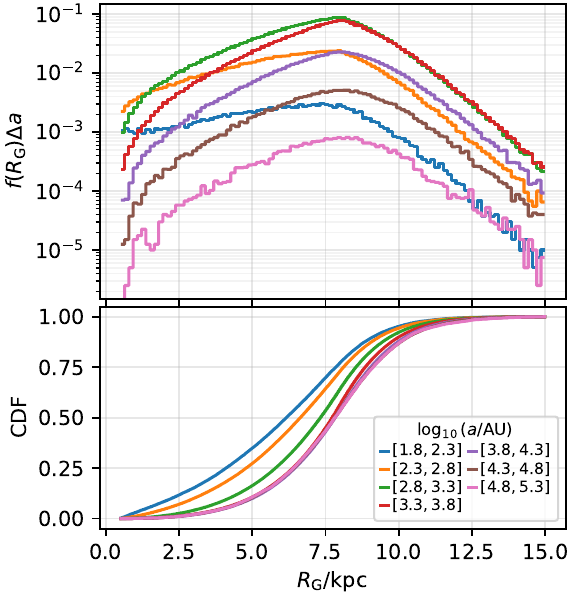}{0.321\textwidth}{(d)}
        \fig{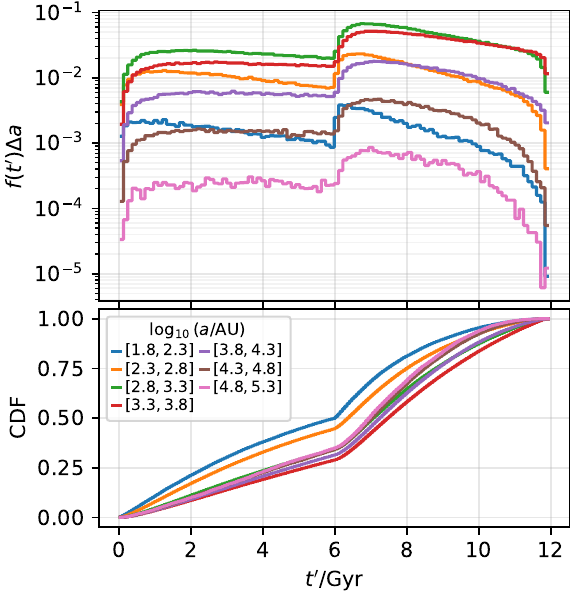}{0.321\textwidth}{(e)}
        }
    \caption{These five subfigures are SMA-focused, ``1.5-dimensional'' versions of the contour plots in Figure~\ref{fig:corner.pdf}, detailing ``where'' and ``when'' 3BBs were assembled before migrating to the solar neighborhood. Color, the extra ``0.5th" dimension, distinguishes between different logarithmically-spaced bins, $i$, in SMA of width $\Delta a_i$. The bottom panel of each subfigure displays the cumulative density functions, $C(X,a_i)$, of the integration variable $X$ specified on the panels' horizontal axes, for each bin in SMA. The top panels of each subfigure display the corresponding probability density functions scaled by SMA bin width, $f(X)\Delta a_{\rm i}=\frac{\partial^2 C(X,a_{\rm i})}{\partial X \partial a} \Delta a_{\rm i}$. The integration variables under examination are (a: $\{r/r_{\rm h}\}$) radial cluster location of 3BBF in terms of the natal cluster's half-mass radius, (b: $\{\tilde{\tau}\}$) time between cluster formation and 3BBF as a fraction of the total cluster lifetime, (c: $\{M_{\rm cl}\}$) mass of the 3BB's natal cluster at time $\tilde{\tau}$, (d: $\{R_G\}$) Galactocentric radius of natal cluster, (e: $\{t'\}$) time the 3BB's natal cluster dissolves, where $t'=12\,\rm Gyrs$ is the present day. The widest binaries tend to be born later in time, both in $\tilde{\tau}$ and $t'$, and further out, both in $r/r_{\rm h}$ and $R_{\rm G}$, than tight binaries. Across the entire parameter space, we predict that solar neighborhood 3BBs mostly formed in clusters with masses $M_{\rm cl}(\tilde{\tau}) < 10^2\,M_\odot$, near the point of absolute dissolution of the cluster ($\tilde{\tau}{\sim}1$).}
    \label{fig:five_plots}
\end{figure*}

\begin{figure*}
    \fig{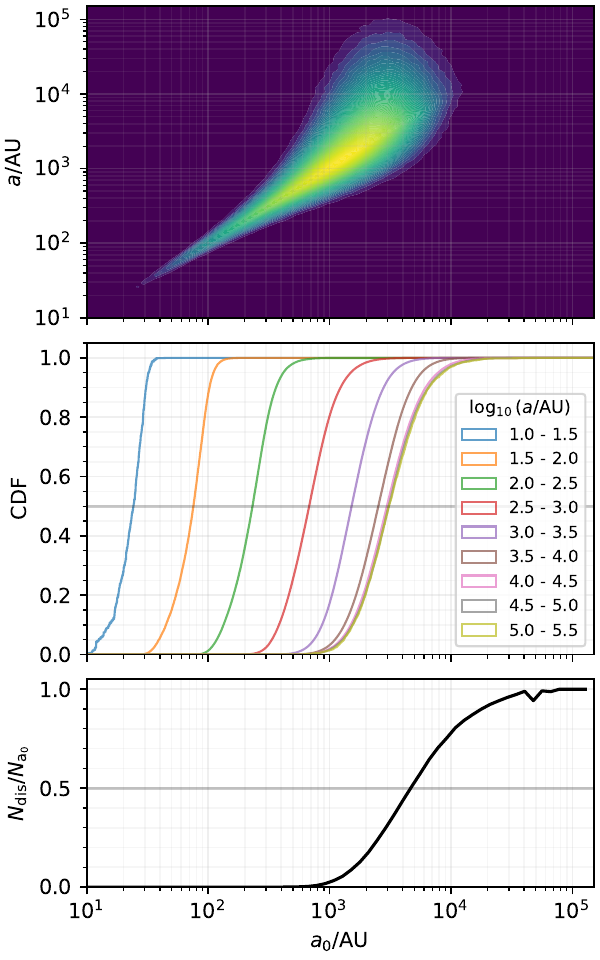}{0.4\textwidth}{(i)}
    \fig{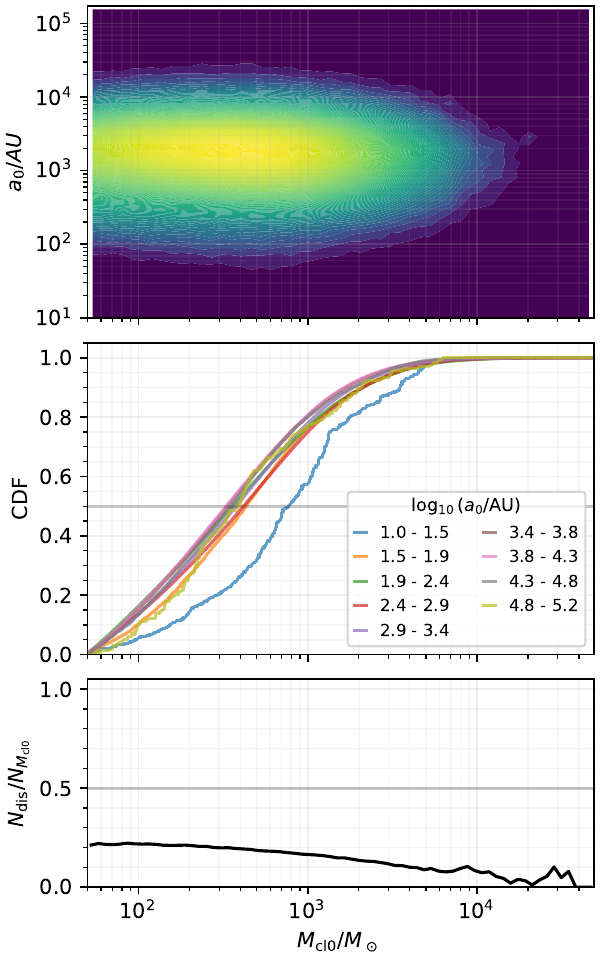}{0.4\textwidth}{(ii)}

    \caption{Contour plots (\textit{top}) and an accompanying set of cumulative distribution functions (\textit{middle}) illustrating the relationship between (i) the initial binary SMA, $a_0$, and its final SMA, $a$, following \textsc{CS} field evolution, and (ii) $a_0$ and the initial mass of the binary's natal cluster, $M_{\rm cl0}$. In column (ii), we include the samples of binaries eventually disrupted through \textsc{CS}. The \textit{bottom} panels show the fraction of disrupted field binaries given (i) $a_0$ or (ii) $M_{\rm cl0}$. The tighter the initial binary, the less its SMA tends to change. Critically, we predict that wide field 3BBs in the solar neighborhood today with SMA $a>4000\,\rm{AU}$ are overwhelmingly born with $a_0<4000\,\rm{AU}$. Hence, the wider the initial binary, the more rapidly it is destroyed in the field. Separately, low-mass clusters tend to produce the highest numbers of binaries, with over $75\%$ of binaries born in clusters with $M_{\rm cl0}<10^3\,M_\odot$.}
    \label{fig:a0a1.pdf}
\end{figure*}

\subsection{Formation Characteristics of Solar Neighborhood Three-body Binaries}\label{sec:3bbf formation characteristics}
Figure~\ref{fig:corner.pdf} broadly highlights the underlying distribution and dimensional covariance found by evaluating the G3R. There are a few intuitive takeaways here: (i) most field 3BBs are created early in a cluster's life when it is densest, according to the united equations of evolution (UEE, \S~\ref{sec:cluster evolution}), with the smallest binaries born almost exclusively in a cluster's infancy, (ii) very few binaries have orbital radii which migrate more than $3 \, \rm{kpc}$ through the disk, as expected from the \citet{Frankel_2020} diffusion model (\S~\ref{sec:Milky Way Evolution}), and (iii) there is a sharp preference for field 3BBs to be sourced from clusters born at the onset of star formation in the thin disk ($t=6$~Gyrs in \citealt{Frankel_2020}).

A more robust physical intuition may be found by examining ``1.5-dimensional'' versions of the contour plots relating binary SMA to cluster and MW history; they are displayed in Figure~\ref{fig:five_plots}. Here, the dimensions (integration variables) of the G3R under scrutiny are (from left to right): $r/r_h$, the radial location within its host cluster where a 3BB is born, $\tilde{\tau}$, the fraction of the cluster's lifespan elapsed when the 3BB forms and escapes, $M_{\rm cl}$, the mass of the cluster at the time of 3BBF and escape, $R_G$, the Galactocentric radius of the birth cluster's circular orbit, and $t'$, the age of the MW when the 3BB was born. Several interesting predictions are discernible here. First, roughly half of 3BBs are born outside the host cluster's core ($r\gtrsim 0. 5r_h$), with wider binaries being born both further from the cluster's center and later in the cluster's life than tighter binaries. Aside from the tightest 3BBs, most are born near dissolution in the final ${\sim}20\%$ of the cluster's life. 

The overwhelming majority of 3BBs originate from initially low-mass clusters ($M_{\rm cl0}<10^3\,M_\odot$) due to their dominance of the CIMF (Equation~\ref{eq:cimf}). Surviving wide binaries are born nearest to cluster dissolution, while tight binary formation occurs throughout a cluster's life. Additionally, our model predicts that the tightest, solar neighborhood 3BBs should be born uniformly through the MW disk from $R_G = 1$--$8\,\rm{kpc}$ and early in MW star formation history, with $50\%$ born within the first $6\,\rm{Gyr}$ of the MW's life (assuming a present MW age of $12\,\rm{Gyr}$). Wider binaries are born later in MW history and nearer to the solar neighborhood, an intuitive requirement due to the profound increase in a binary's ionization cross section with an increasing binary SMA (\S~\ref{sec:disk disruption}).

Dissolving clusters have long been postulated to be reliable environments for dynamical wide binary formation \citep{Sterzik_1998, Moeckel_2010, Kouwenhoven_2010, Moeckel_2011}. We corroborate this, as we find isolated 3BBF to be a significant mechanism for generating solar neighborhood binaries when surviving till cluster dissolution $(\tilde{\tau}\approx1)$. Admittedly, our idealized treatment of star clusters as single-component Plummer models evolving under the UEE (\S~\ref{sec:cluster evolution}) is unlikely to reliably model cluster dissolution. The conditions for \textit{isolated} 3BBF may also be rare in a cluster with few (${\lesssim}50$) bodies, though there is evidence that binary formation in strong interactions involving \textit{more} than three bodies at the center of a cluster could be a significant source of tight binaries \citep{Tanikawa2013}. Such binaries are formed during particularly strong spikes in the cluster's central density. Even our idealized cluster model exhibits a density spike at dissolution, so a more nuanced investigation of binary formation near dissolution may be required to better assess the validity of our assumptions surrounding 3BBF at dissolution.

The left-hand panels of Figure~\ref{fig:a0a1.pdf} draw a relationship between the SMA before field evolution (pre-CS) and the final SMA of a 3BB population within the solar neighborhood today. As expected, the tighter the initial binary, the less \textsc{CS} shifts its SMA. A critical, population-level transition presents itself for the widest binaries: 3BBs with final SMAs ${>}4000\,\rm{AU}$ overwhelmingly escaped their natal clusters with initial SMAs ${<}4000\,\rm{AU}$. In other words, field evolution preferentially destroys the widest field 3BBs, and widens initially tighter binaries $({\sim}10^3$--$10^{3.5}\,\rm{AU})$.  If we were to turn off disruption during \textsc{CS} field evolution, the distribution of wide binary SMA would become egregiously skewed to very wide SMA. The dearth of such very-wide binaries in our models is simply due to disruption during field evolution. Thus, any prospective wide binary formation channel that matches the observed Gaia SMA distribution, while neglecting field evolution, cannot account for most of the Gaia wide binary population.

Finally, the relationship between initial binary SMA and the initial mass of the natal cluster is displayed in the right-hand panels of Figure~\ref{fig:a0a1.pdf}. These distributions include the contribution of binaries which would ultimately be disrupted through \textsc{CS}, the left panel does not. Our models reinforce the general consensus that the field binary generation is dominated by low-mass clusters \citep{Moeckel_2011}.

\section{Discussion \& Summary}
\label{sec:summary}

\subsection{Comparison to Alternative Binary-Formation Channels}

Tight field binaries have been historically assumed to form primordially \citep{King_2012, Bate_2014, Parker_2014}, the remnants of which are the select few binaries that have avoided disruption. Recent state-of-the-art simulations from the \textsc{Starforge} collaboration do in fact find that many binaries form primordially in clusters \citep{Guszejnov_2023, Farias_2024}, with a primordial binary fraction of ${\sim}40\%$. The majority are bound to the cluster with very wide SMAs and are unlikely to survive the ensuing dynamics in such a dense stellar environment. However, 3BBF may help serve as a complementary formation channel for tight subthermal binaries, the overwhelming majority of which originate almost exclusively in the earliest, densest stage of cluster history within our models. While we do not assert 3BBF as a prominent method for tight binary generation, primordial binary formation need not be the only means.

Wide Gaia binary formation channels have been more elusive, primarily because the proposed formation channels readily produce many wide binaries---e.g., cluster dissolution or random pairing \citep{Kouwenhoven_2010, Moeckel_2010,Penarrubia2021}---or a few highly eccentric binaries---e.g., disk fragmentation \citep{Leeetal2017, Xu_2023}---but not necessarily both in the required quantities. In particular, before this work, there had not yet been a thorough investigation of the cluster formation and dissolution channel's contribution to the present-day field binary population. As highlighted in \S~\ref{sec:3bbf formation characteristics}, the tendency for binary field evolution is to accelerate the widening of binaries towards disruption. This implies that any prospective wide binary formation channel matching the observed Gaia distribution, prior to the application of field evolution, cannot be the dominant source of the Gaia wide binary population.

With 3BBF, the formation of highly eccentric wide binaries is preferred, especially late in a cluster's life---supporting the cluster dissolution channel. By accounting for the disruptive capacity of dynamical encounters in clusters and galactic field perturbations, we find that 3BBF may contribute a significant fraction of Galactic wide binaries observable today.

\subsection{Limitations and Future Work}

For a first foray into assembling a Galactic 3BBF rate, we opted for a simple, physically motivated astrophysical model. That said, there are several practical developments that can further improve the physical realism and self-consistency of our method.  

Substantial effort is already devoted to modeling binary destruction in the natal cluster (\S~\ref{sec:escape_disruption_cluster}) and in the Galactic field (\S\S~\ref{sec:disk disruption} and \ref{sec:cumulative scatter}). What we do not follow is the effect of in-cluster dynamical interactions \textit{not} fully disrupting a 3BB, instead opting to forcefully discount any binary that undergoing even a weak encounter. These dynamical interactions will typically widen low-mass, wide binaries and further thermalize their eccentricities \citep{Heggie_1975}. Ignoring them does not alter our conclusions: the few wide binaries not disrupted in-cluster would provide a subtle thermalization that would only push the 3BB eccentricity distributions in Figure~\ref{fig:gaia_ecc_all.pdf} towards the more thermal Gaia eccentricities identified by \citet{Hwang_2022}. Additionally, while we incorporate a simple giant molecular cloud disruption recipe (\S~\ref{sec:cumulative scatter}), we do not consider the possibility of disruption from other MW substructures, like dark matter subhalos or other star clusters. Incorporating said structures alongside a time-dependent density profile of the MW would help improve the physical accuracy of our framework and increase the destruction rate of 3BBs in the Galactic field.

Tidal circularization, orbital decay, and common-envelope episodes from isolated binary stellar evolution (BSE), has also been neglected. BSE is especially important for the extremely eccentric binaries readily produced by 3BBF \citep{Stegman_2024}. Predicting the stellar metallically of binary members also requires an accurate treatment of BSE, which we leave for future work.

Combining the cluster evolution equations from \citet{Gieles_2011} and an equal-mass Plummer model is a fast and transparent means to rapidly evaluate cluster properties across mass scales. However, a more refined prescription for the structure and evolution of star clusters with realistic initial mass functions and stellar populations will greatly improve the physical realism of our model. Notably, black holes (BHs) play critical roles in the evolution of star clusters  \citep{Merritt_2004,Mackey2007,Mackey2008, BreenHeggie2013, Morscheretal2015,Peuten2016,Chatterjee2017a,Kremer2019,Weatherford_2020,Weatherford2021,Gieles2021,Gieles2023,Roberts2025}. After segregating to the cluster core, their high masses, low velocities, and high local density make BHs the primary catalyst for assembling compact object 3BBs \citep[e.g.,][]{Ivanovaetal2005, Morscheretal2015, Weatherfordetal2023}---and possibly for wider 3BBF more generally.

It is unlikely compact objects comprise many of the wide, eccentric binaries which are the focus of this work. As discussed in \S~\ref{sec:3bbf formation characteristics}, wide superthermal 3BBs that survive and successfully migrate to the solar neighborhood overwhelmingly originate outside of the core of low-mass clusters $(M_{\rm cl0}\lesssim10^3 M_\odot)$ and are catalyzed within the final ${\sim}20\%$ of the cluster's life. These clusters would host only a few compact objects, which primarily reside in the cluster core. We therefore expect the surviving wide 3BBs to be mostly low-mass stellar pairs. Additionally, the interaction cross-section of wide binaries born outside the cluster's core is dominated by single stars, encounters our framework already accounts for conservatively (see \S~\ref{sec:escape_disruption_cluster}). Thus, the key benefit to incorporating a richer stellar population would be to uncover subtle pairing tendencies between different stellar types.

Modeling 3BBF in multi-mass clusters will require a semi-analytic treatment of unequal mass 3BBF. The first such study by \citet{Atallah_2024} determined that \textit{positive} energy 3UB encounters generally favor the pairing of the two least massive bodies. This revised the prior notion that the two most massive bodies are the most likely to bind \citep[e.g.,][]{Morscheretal2015}, which is the case for three-body encounters with \textit{negative} total energy, such as binary--single interactions \citep{StoneLeigh2019,GinatPerets2021,Kol2021}. Unequal-mass 3BBF produces only modest shifts in SMA and eccentricity distributions when mass ratios vary by  ${\lesssim} 2$, so the overall Gaia-binary demographics should not change dramatically. The largest impact should appear in hard, compact-object–plus-star binaries, whose extreme mass ratios demand an explicit unequal-mass treatment. The required parameter space is vast and subtle, calling for a dedicated study and, ideally, full cluster $N$-body simulations—an effort we plan to undertake. 

Our models are likely insufficient at characterizing binary formation late into cluster dissolution, whether through 3BBF or more complicated scattering modes. \citet{Kouwenhoven_2010} estimated a wide binary fraction present at cluster dissolution of $10\%$ to $20\%$. This corresponds to one to two wide binaries formed per cluster at dissolution, though \citet{Moeckel_2011} found integrating their models forward in time by a factor of two diminished the binary survival rate by an order of magnitude, leaving only a fraction of clusters emitting a single binary at dissolution. This finding is consistent with our own despite approximating cluster evolution with a simplistic analytic model. Whether the binaries form in the final moments of the cluster's life, or somewhat earlier, through isolated 3BBF is left for a future investigation.

Finally, our model already closely matches or exceeds Gaia's DR3 binary counts despite applying several conservative filters for binary survival. More complicated channels tied to cluster dissolution (e.g., \citet{Kouwenhoven_2010, Moeckel_2011}) will overlap with 3BBF because realistic dissolution makes no such distinction between dynamical formation mechanisms.  Any concern surrounding over-counting should be allayed by the eventual inclusion of multi-mass 3UB scattering. The 3BBF rate decreases as the mass ratio between the most- and least-massive scattering bodies diverges from unity \citep{Atallah_2024}, favoring 3BBF with comparable masses, thus reducing the overall 3BBF rate.  Additionally, the current Gaia binary counts are likely an underestimate \citep{Elbadry_2018, ElBadry_2021}, so we expect the catalog of wide binaries to expand in Gaia DR4, potentially pushing Gaia observations and our theoretical results into greater agreement.

\subsection{Summary}
By developing a semi-analytic Galactic three-body binary formation rate (G3R), we have demonstrated that three-body binary formation (3BBF) can efficiently generate highly eccentric, wide field binaries. Our model predicts that binaries with SMA ${<}10^3 \rm AU$ exhibit subthermal eccentricities whereas binaries with SMA ${>}10^3 \rm\,AU$ are superthermal. Contingent on physically-motivated assumptions regarding cluster-structure and tides, this framework closely matches the observed SMA distribution of Gaia's wide binaries \citep{ElBadry_2021} and the eccentricity distributions of \citet{Hwang_2022} when applying dynamical binary field evolution framework constructed by \citet{Hamilton_2024}. The predicted number of wide binaries in the solar neighborhood matches or exceeds that observed by Gaia, though we have not accounted for observational incompleteness or selection effects. However, these systematics primarily affect Gaia's ability to observe binaries with separations $<200 \rm AU$. Thus, 3BBF may be an essential formation channel for the widest field binaries.

Future investigations incorporating unequal-mass 3BBF, binary stellar evolution, unequal-mass clusters, more advanced cluster evolution, and improved Milky Way disk evolution models will greatly enhance the accuracy and realism of our framework, as will accounting for Gaia selection effects. If originating heavily from 3BBF, Gaia's wide binaries are artifacts of the internal dynamics in our Galaxy's dense stellar environments---especially those that may have dissolved by the present---and of the Galaxy's evolutionary history. It immediately follows that Gaia wide binaries may be powerful probes of dissolved star clusters and of Galactic archaeology more broadly.

\begin{acknowledgments}
We are grateful for insightful conversations shared with J.~Binney, N.~Choksi, N.~Frankel, M.~Gieles, C.~Johnson, B.~Kocsis, K.~Kremer, J.~Li, A.~Marszewski, J.~Magorrian, C.~O'Connor, T.~Panamarev, H.B.~Perets, S.~Rose, M.~Rozner, B.~Scott, and N.C.~Stone on topics ranging from cluster formation history to Milky Way evolution and dynamics. D.A.~would like to thank Fred Rasio for his commitment and support towards the production of this work and the author's doctorate. D.A.~acknowledges support from the CIERA Board of Visitors Fellowship. This work was supported in part by a Leverhulme Trust International Professorship Grant (No.~LIP-2020-014). The work of Y.B.G.~was partly supported by a Simons Investigator Award to A.A.~Schekochihin. This work used computing resources at CIERA funded by NSF PHY-2406802. This research was supported in part through the computational resources and staff contributions provided for the Quest high performance computing facility at Northwestern University which is jointly supported by the Office of the Provost, the Office for Research, and Northwestern University Information Technology.

\end{acknowledgments}


\vspace{5mm}
%\facilities{HST(STIS), Swift(XRT and UVOT), AAVSO, CTIO:1.3m, CTIO:1.5m,CXO}




\appendix

\section{Modeling Three-body Binary Formation in Star Clusters}
\label{sec:Modeling Three-body Binary Formation in Star Clusters}

\subsection{The 3BBF Rate}\label{sec:The 3BBF Rate}
From \citet{Atallah_2024}, the volumetric rate of encounters between \textit{three unbound bodies} (3UBs) for a field of equal masses, $m$, with a Maxwellian velocity distribution is
\begin{equation}\label{eq:equalmassencrate}
\tilde{\Gamma}_{3B} = \frac{d \Gamma_{3B}}{d \volume} = \frac{2^{5/2} \pi^{3/2}}{3} n^3 R_1^5 \sigma,
\end{equation}
where $n$ is the local number density, $\volume$ denotes volume, $\sigma$ is the local velocity dispersion, and $R_1$ is the radius of the largest spherical volume (henceforth the ``interaction volume'') containing all three bodies during the encounter if those bodies were to follow straight-line trajectories.

The 3BBF rate is simply the 3UB encounter rate multiplied by the numerically determined probability of binary formation, $P_F$, in 3UB encounters with the given $n$, $R_1$, and $\sigma$. \citet{Atallah_2024} calculated this semi-analytic 3BBF rate with similar initial scattering conditions to \citet{AH76}, setting $R_2 = 15 R_1$. Here, $R_2$ is the offset distance between the slowest moving body and the interaction volume. The other two bodies are initiated further from $R_1$, at distances proportional to their randomly drawn initial velocity, assuming a Maxwellian velocity distribution with dispersion $\sigma$.\footnote{See \citet{Atallah_2024} for a more thorough numerical exploration of the simultaneous scattering of three unbound bodies (3UB).}  In a dense cluster environment, if $R_1 \sim r_{\rm sep}$, the average local interparticle distance, then it does not make sense for bodies to travel a distance $R_2 = 15 R_1$. The individual bodies would likely have a strong encounter en route to the 3UB interaction, breaking the approximation of an isolated encounter. The equal-mass 3BBF scattering probability of \citet{Atallah_2024} thus needs to be refined by reducing $R_2/R_1$ from $15 $ to $3$. The resulting numerically determined 3BBF probability is
\begin{equation}\label{eq:PF}
\begin{aligned}
P_F &= c_0 \left(\frac{2 b_{90}}{R_1}\right)^2 = c_0 \left(\frac{2 G m}{3 \sigma^2 R_1}\right)^2, \ c_0=3.82,
\end{aligned}
\end{equation}
where $b_{\rm 90} = G m/(3 \sigma^2)$ is the impact parameter in a gravitational two-body encounter that deflects the bodies by $90^\circ$ from their original trajectory. The same scattering and fitting method from \citet{Atallah_2024} is used, limiting the domain to values of $R_1/(2 b_{90})>10$ (see \S~3.1 and Figure~4 in \citet{Atallah_2024} for more information).

Using our equal-mass formation probability, the resulting volumetric formation rate is
\begin{equation}\label{eq:gammaf0}
    \begin{aligned}
    \tilde{\Gamma}_{F,0} = P_F \tilde{\Gamma}_{3B} &= c_0 \frac{2^{9/2}
    \pi^{3/2}}{3^3} \frac{G^2 m^2}{\sigma^3} n^3 R_1^3\\
    &= c_0 \frac{2^{5/2}
    \pi^{3/2}}{3^2} \frac{G^2 m^2}{\sigma^3} n^2 k^3.
    \end{aligned}
\end{equation}
For convenience, $R_1$ is rewritten in terms of the average interparticle separation, $r_{\rm sep} = \left(4 \pi n/3\right)^{-1/3}$, becoming $R_1 = k\, r_{\rm sep}$. The $G^2 m^2 n^2 \sigma^{-3}$ scaling above is notably shallower than the classic $G^5 m^5 n^3 \sigma^{-9}$ scaling found in detailed balance calculations \citep{Heggie_1975, Goodman_1993} and recovered recently in our modern approaches \citep{Ginat_2024, Atallah_2024}, which applies to the formation of \emph{hard} binaries. Prior estimates fixate on hard 3BBF, whereas we enable both locally hard and soft 3BBF. By relaxing the limit on binding energy and, instead, enforcing a geometric limit based on particle separations, we recover a formation rate scaling far more enabling of 3BBF, albeit with arbitrarily small binding energies. In the limit of hard binary formation, the combination of Equations~(\ref{eq:gammaf0}) \& (\ref{eq:fae}, see \S~\ref{sec:The SMA/Eccentricity Distribution}) yields the classic scaling. 

The validity of the above equations are contingent upon $n \volume \ll 1$. To generalize Equation~(\ref{eq:equalmassencrate}), and so allow the consideration of encounters with $n\volume \simeq 1$, the expression
\begin{equation}\label{eq:Piso1}
    P_{\rm iso} d\lambda = e^{-\lambda} d\lambda, \ \lambda=n \volume
\end{equation}
must be applied; $P_{\rm iso}$ is the distribution of isolated encounters (e.g., no other bodies present within $R_1$) under the assumption that the rate of encounters may be modeled as a Poisson process. Thus,
\begin{equation}\label{eq:Piso2}
    P_{\rm iso} d\lambda = 3 k^2 e^{-k^3} dk
\end{equation}
where $k$ is the radius of the interaction volume in units of $r_{\rm sep}$.

It is straightforward to integrate over all $k$, but, realistically, tidal forces from the cluster will tear apart an encounter before it occurs if $k$ is large enough. To account for this, a transformation can be applied to values of $k$ appearing in Equation~(\ref{eq:gammaf0}), limiting the maximum possible interaction volume self-consistently. A convenient choice for the maximum value of $k$ is some fraction, $\eta_{\rm H}$, of the Hill Radius, $R_H \approx \, r \left(\langle m\rangle /M_{\rm cl,enc}\right)^{1/3}$; $M_{\rm cl,enc}$ is the total cluster mass enclosed at the radial location of the encounter. In terms of $r_{\rm sep}$, the fraction of local Hill Radius in a Plummer cluster is $k_H=\eta_{H} (1+(r/b)^2)^{-1/3}$, where $\eta_H$ is arbitrarily selected to be $\ll 1$; we explore the effects of conservatively limiting $\eta_H$ in our results (\S~\ref{sec:results}). The transformation applied is thus: $k\rightarrow k_1=\min(k, k_H)$. Specified to the case of encounters within a single field, and including the Hill Radius limit, Equation~(\ref{eq:gammaf0}) becomes\footnote{Applying the $k_1$ transformation is equivalent to evaluating the integral $\int_0^k \tilde{\Gamma}_{F,0}(k_1)P_{\rm iso}(k)dk = \int_0^{k_H} \tilde{\Gamma}_{F,0}(k)P_{\rm iso}(k)dk + \tilde{\Gamma}_{F,0}(k_H) \int_{k_H}^k P_{\rm iso}(k) dk$. The transformation should not be applied to Equation~(\ref{eq:Piso2}) since it is only the distribution of all possible isolated encounters and should integrate to 1 over all $k-$space independent of the local physics in an environment. In the limit that $k_H^3 \ll 1$, the integral reduces to $\int_0^k \tilde{\Gamma}_{F,0}(k_1)P_{\rm iso}(k)dk\approx \tilde{\Gamma}_{F,0}(k_H)$.}
\begin{equation}
    \begin{aligned}
        \tilde{\Gamma}_{F} &= P_{\rm iso} \tilde{\Gamma}_{F,0} = c_0 \frac{2^{5/2} \pi^{3/2}}{3} \frac{G^2 m^2 n^2}{\sigma^3} k_1^3\,  k^2 e^{-k^3}.
    \end{aligned}
\end{equation}

The above expression represents the net 3BBF rate across all possible $\{a,\epsilon\}$. In the next section, a 2D distribution function is assembled, allowing the 3BBF rate to be modeled as a function of $\{a,\epsilon\}$.

\subsection{The SMA/Eccentricity Distribution}\label{sec:The SMA/Eccentricity Distribution}

\begin{figure}
    % First PDF
    
    \gridline{\includegraphics[width=.48\columnwidth]{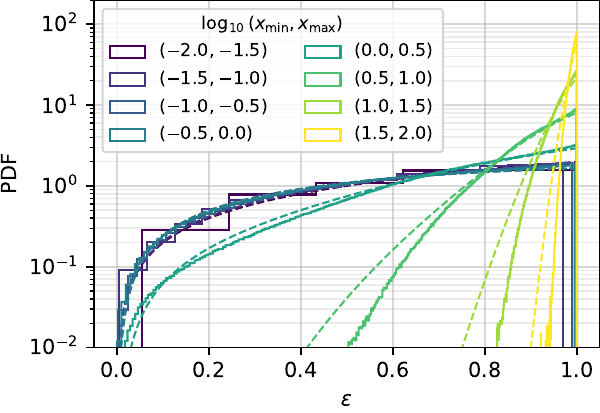}
    % Second PDF
    \includegraphics[width=.48\columnwidth]{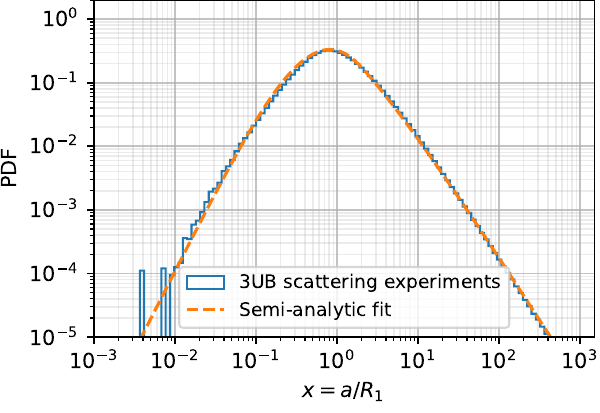}}
    \caption{Eccentricity ($\epsilon$) and semi-major axis ($a$) distributions for 3BBF from the scattering experiments described in \S~\ref{sec:The 3BBF Rate}. Note we normalize $a$ by $R_1$, the radius of the \textit{interaction volume}---the largest spherical volume containing all three bodies during the encounter if those bodies were to follow straight-line trajectories. The solid lines are the raw results of the scattering experiments while the dashed lines show the distributions as approximated by Equation~(\ref{eq:fae}). \textit{Top}: Eccentricity distribution for several $x=a/R_1$ bins. As found by \citet{Ginat_2024} and \citet{Atallah_2024}, binaries with $a>R_1$ have 
    $\epsilon$ distributions that become increasingly superthermal with increasing $a$, but are thermal when $a<R_1$; the $\epsilon$ distribution is actually subthermal for $x\in[10^{-0.75},10^{0.0}]$, but this is not to resolved above. \textit{Bottom}: Normalized semi-major axis distribution. As predicted by \citet{Heggie_1975, Goodman_1993}, $f(x) \propto x^{5/2}$ when $a \ll R_1$,  though the latter quoted the distribution cumulatively as $C(x)\propto x^{7/2}$.}
    \label{fig:3BBF_ae_distribtuions}
\end{figure}

Prior to \cite{Ginat_2024,Atallah_2024}, the joint SMA ($a$) and eccentricity ($\epsilon$) distribution of binaries formed exclusively through 3BBF were not known \citep{AH76,Goodman_1993,Ivanovaetal2005}. \cite{Ginat_2024,Atallah_2024} found that 3BBF features a constant, thermal distribution ($f_F(\epsilon)=2\epsilon$; see \citealt{Jeans_1919}) for binaries with an SMA $a < R_1$ and an increasingly superthermal distribution for all SMA in the regime $a>R_1$. There is a slight subthermal region that peaks between $10^{-0.75}$ and $1$. This subthermal region plays an outsized role in the eccentricity of tight binaries because the formation rate of binaries with $a<R_1$ peaks when $a\approx R_1$. Remarkably, the 2D $(\epsilon,a/R_1)$ distribution exhibits (near-) scale-free behavior for regions of the parameter space corresponding to $\chi_1=R_1/(2 b_{90}) \gtrsim 10$. Thus, independent of the 3BBF probability, $P_F(\chi_1>10)$, the 2D probability density function $f(\epsilon, a/R_1) \, d\epsilon \, da/R_1$ may be semi-analytically approximated with ease.

Employing the \textsc{PySR} symbolic-regression tool \citep{PySR} and feeding it ${\sim}10^9$ $\{\epsilon,a/R_1\}$ samples from the new set of 3UB scatterings with $R_2 = 3 R_1$ (generated to calculate Equation~(\ref{eq:PF})), the 2D PDF between $a$ and $\epsilon$ in the equal-mass, equal-velocity distribution limit may be approximated as
\begin{equation}\label{eq:fae}
\begin{aligned}
    f_F(\epsilon,x) \, d\epsilon \, dx &= \frac{1.98965 \, \epsilon^{x} \, d\epsilon \, dx}{\epsilon^{3-2x} + \frac{0.088857}{\epsilon \, x^{2.5}} + \frac{1.25678}{x}+x^\epsilon}\\  
    &x = a/R_1, \ dx = da/R_1, R_1 = k_1 r_{\rm sep},
\end{aligned}
\end{equation}
where we have employed the Hill Radius correction in $k_1$ (\S~\ref{sec:The 3BBF Rate}) and is normalized over the domain $\{ {\epsilon \in [0,1]}, \ {x\in(0,\infty)} \}$. The $a$ and $\epsilon$ distributions from our 3UB simulation data and semi-analytic approximation are displayed in Figure~\ref{fig:3BBF_ae_distribtuions}. Like \citet{Goodman_1993}, we find that the hardest/closest binaries ($a\ll R_1$) exhibit a 3BBF SMA distribution $\propto a^{5/2}$. This SMA scaling just so happens to be identical to the observed close-binaries in the Gaia binary curve (\S~\ref{sec:Gaia sma and ecc}).  Updating the 3BBF formation rate by incorporating a dependence on $(\epsilon, a)$ and $k$, it may be written
\begin{equation}\label{eq:gamma_ex}
    \frac{d^3 \tilde{\Gamma}}{dk \, d\epsilon \, dx} = \tilde{\Gamma}_{F}  f_F(\epsilon,x).
\end{equation}

\subsection{3BBF in a Plummer Cluster}\label{sec:Plummer model}

Our focus is to estimate the rate of 3BBF in \textit{stellar clusters}. As we introduce nine integration variables, the calculation will quickly become intractable unless we employ a simple analytic cluster model, such as a Plummer model, to estimate local cluster properties. For future reference, some useful Plummer expressions include

\begin{equation}\label{eq:plum_eqns}
    \begin{aligned}
        &\rho = \frac{3 M_{\rm cl}}{4 \pi b^3 \left(1 + (r/b)^2 \right)^{5/2}}, \ \Phi = -\frac{G M_{\rm cl}}{b \left(1+(r/b)^2\right)^{1/2}}, \ \sigma^2 = -\frac{\Phi}{6}, \ v_{\rm esc} = \sqrt{12} \sigma, \ \
        b=  (2^{2/3}-1)^{1/2} r_{\rm h}  
    \end{aligned}
\end{equation}
where $\rho=\langle m \rangle n$, $n$, $\Phi$, $\sigma$, $M_{\rm cl}$, $\langle m \rangle$, $b$, and $r$ are the local mass density, number density, potential energy, one-dimensional velocity dispersion, total cluster mass, average mass, Plummer scale-length (or kernel), and radial distance from the center of the Plummer cluster, respectively.

Equation~(\ref{eq:gamma_ex}) may be written in terms of the Plummer relations, becoming
\begin{equation}\label{eq:gamma_plum_ex}
\begin{aligned}
&\frac{d^3 \tilde{\Gamma}}{dk \, d\epsilon \, dx} = \frac{d^4 \Gamma}{d\volume dk \, d\epsilon \, dx} = \frac{d^4 \Gamma}{4 \pi r^2 dr \, dk \, d\epsilon \, da/R_1},\\
&\frac{d^4 \Gamma}{dr dk d\epsilon da} =  36 c_0 \sqrt{\frac{3 G \langle m\rangle}{\pi b^7}} \left(\frac{M_{\rm cl}}{\langle m\rangle } \right)^{5/6} k_1^2 k^2 e^{-k^3}  \frac{(r/b)^2}{(1 + (r/b)^2)^{61/12}}f_F\left(\epsilon,\frac{a}{k_1 r_{\rm sep}(r)}\right).
\end{aligned}
\end{equation}
We have written the 3BBF rate in terms of the radial displacement in the Plummer cluster, $r$, and SMA, $a$, instead of $\volume$ and $x=a/R_1$ for integration convenience. In the final portion of the calculation (\S~\ref{sec:G3R}), the substitution $r\rightarrow r_t \tilde{r}$ is applied to limit the integration boundary within the tidal radius of the Plummer cluster; $\tilde{r}$ is the radial displacement normalized by the tidal radius of the cluster. Equation~(\ref{eq:gamma_plum_ex}) is the ``local creation rate'' and we will not consider any other means to form binaries. The ``creation rate'' is now written 
\begin{equation}
    \frac{d^4 \Gamma}{dr dk d\epsilon da} \equiv \mathcal{C}(M_{\rm cl}, b, r,k,\epsilon, a)
\end{equation}
for {\ae}sthetic compactness. Subsequent sections will assemble probability distributions describing the evolution of individual clusters and the Milky Way or binary disruption in the cluster or the field .

\subsection{Cluster Evolution}\label{sec:cluster evolution}
Emphasizing practicality over absolute accuracy, we adopt the \citet[Appendix C]{Gieles_2011} Unified Equations of Evolution (UEE) cluster model for its simplicity and natural compatibility with the entirety of the nine-dimensional distribution function under assembly. While these models are not intended to provide exact predictions, they are a convenient approximation to apply across a wide range of cluster masses, $M_{\rm cl}$, and Galactocentric orbital radii, $R_G$. Comparing the analytic timescales of Appendix A, UEE in Appendix C, and Table 2 (all found in \citet{Gieles_2011}), the UEE become
\begin{equation}\label{eq:UEE}
\begin{aligned}
    &M_{\rm cl}(M_{\rm cl0},\tilde{\tau}) = M_{\rm cl0} \left(1 - \tilde{\tau} \right)^{4/3}, \ M_{\rm cl0}= \langle m\rangle N_{\rm cl0}, \ \tilde{\tau}=\tau/\tau_{\rm cl}, \\
    &\tau_{\rm cl}(N_{\rm cl0}, R_G)= 0.13 N_{\rm  cl0}^{3/4}\frac{R_G}{V_G},\\
    &r_{\rm h}(M_{\rm cl}, R_G) = 0.66 \zeta^{2/3} r_t(M_{\rm cl}, R_G)\, \frac{\ln\left[\Lambda(N_{\rm cl})\right]^{2/3}}{N_{\rm cl}^{1/6}} \left( 1-\left(1-\tilde{\tau}\right)^{11/3}\right)^{2/3},\\
    &r_t(M_{\rm cl}, R_G)= \sqrt[3]{\frac{G M_{\rm cl}(\tilde{\tau})}{2}\left(\frac{R_G}{V_G}\right)^2},
\end{aligned}
\end{equation}
where $M_{\rm cl0}$ is the initial cluster mass, $\langle m \rangle$ is the average cluster mass, $\tau$ is the amount of time elapsed since the cluster expelled most of its gas following the onset of star formation, $\tau_{\rm cl}$ is the predicted lifetime of a cluster, $M_{\rm cl}$ and $N_{\rm cl}$ are the cluster mass and cluster particle number at time $\tilde{\tau}$, $R_G$ and $V_G \equiv V_G(R_G)$ are the Galactocentric radius and Milky Way circular velocity at $R_G$ (see \S~\ref{sec:Milky Way Evolution}), $\Lambda(N_{cl})$ is the argument of the Coulomb Logarithm, $r_t(M_{\rm cl}, R_G)$ is the tidal radius of the cluster, and $\zeta=0.15$ is the cluster heating constant.\footnote{We set $\zeta=0.15$ instead of the prescribed $\zeta=0.2$ in  \citet{Gieles_2011} to better match observations of local MW stellar clusters.}   Additionally, our expression for $r_h$ is fixed to $r_h(\tilde{\tau}=0.05)$ for all times $\tilde{\tau}<0.05$ due to the unrealistically high initial cluster densities predicted by the UEE.

The argument of the Coulomb Logarithm, $\Lambda(N)$, is an object of frequent debate, often approximated for small subsets of the $N$-Body parameter space. Traditionally, $\Lambda$ is given the form $\Lambda=\frac{b_{\rm max}}{b_{\rm min}}$, where $b_{\rm max}$ and $b_{\rm min}$ are the maximum and minimum impact parameters considered in scattering encounters between individual bodies in an $N$-Body system \citep{BT_2008}. \citet{Giersz_1996} prescribes a value of $\Lambda\approx0.02 N_{\rm cl}$ for $N_{\rm cl}\sim10^5$, though they find that the argument should be much larger for smaller values of $N$, becoming $\Lambda\approx0.1 N_{\rm cl}$ if $N_{\rm cl}\sim 250$. A convenient analytic expression which qualitatively and quantitatively bridges the extremes of the parameter space falls out of  setting $b_{\rm max} = r_{\rm sep}(r_h)$ and $b_{\rm min}=b_{90}(r_h)$; $r_{\rm sep}(r_h)$ is the average inter-particle distance and $b_{\rm 90}(r_h)=\frac{G m}{3 \sigma(r_h)^2}$ is the $90$-degree deflection angle between equal masses living at the cluster half-mass radius in a single-component Plummer model. The resulting functional form is
\begin{equation}
    \Lambda(N_{\rm cl}) = \frac{N_{\rm cl}^{2/3}}{2} \left(1 - 2^{-2/3}\right)^{-1/3}\approx 0.7 N_{\rm cl}^{2/3},
\end{equation}
which we have adopted across all cluster masses.

\begin{figure*}
\centering
    % First PDF
    \includegraphics[width=.7\columnwidth]{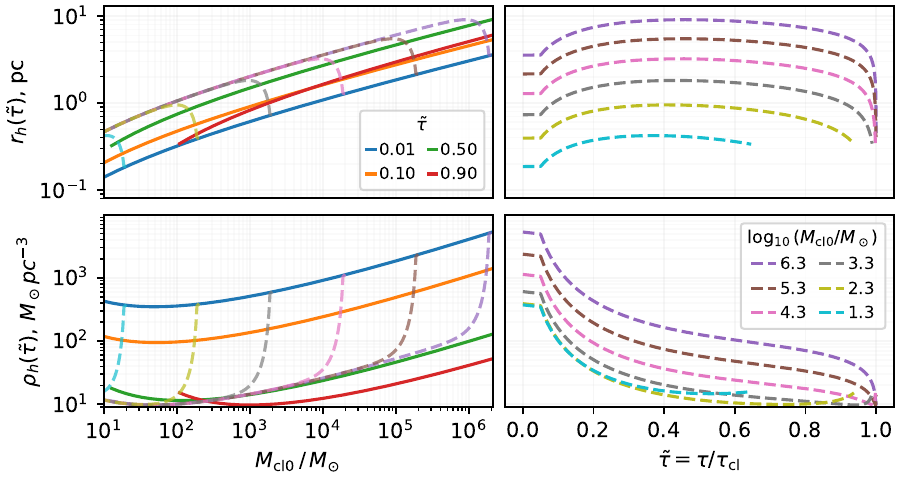}
    \caption{Assumed semi-analytic relations between cluster properties---initial mass $M_{\rm cl0}$, half-mass radius $r_h$, and density $\rho_h$ within the half-mass radius, and age $\tilde{\tau}$ normalized by cluster lifespan---according to our compilation of the UEE from \citet{Gieles_2011}---Equations~(\ref{eq:UEE}). In the left column, the solid curves show the  $r_h$--$M_{\rm cl0}$ relation (\textit{top}) and $\rho_h$--$M_{\rm cl0}$ relation (\textit{bottom}) at several cluster ages indicated by color. The time evolution of $r_h$, $\rho_h$ and $M_{\rm cl}$ for several specific choices of initial cluster mass $M_{\rm cl0}$ are indicated by the dashed curves in both the left and right columns. The early truncation of the dashed yellow and light blue curves prior to $\tilde{\tau}=1$ (cluster dissolution) reflect our choice to halt cluster evolution once there are fewer than 10 bodies, below which these semi-analytic approximations break down. All clusters shown are assumed to evolve on a circular orbit about the Milky Way center at an equivalent Galactocentric radius to the sun ($R_G=8\,\rm{kpc}$).}
    \label{fig:UEE_cluster_evolution}
\end{figure*}

Figure~\ref{fig:UEE_cluster_evolution} displays a host of cluster properties---born on circular orbits at $R_G=8 \, \rm{kpc}$---as they evolve through their cluster lifetime, $\tilde{\tau}=\tau/\tau_{\rm cl}$ and fully dissolved when $\tilde{\tau}=1$. For example, a cluster born with $M_{\rm cl0}=10^{3.3} \, M_\odot$ will have a half-mass radius of $r_h\approx 1.5\, \rm{pc}$ after shedding half its initial mass through relaxation and Galactic tides. All the cluster properties are in qualitative agreement with the set of cluster $M_{\rm cl}, r_{\rm h}$ observations compiled in \citet{Krumholz_2019}.

\subsection{Survival \& Disruption in a Stellar Cluster}\label{sec:escape_disruption_cluster}

Two criteria are required for a 3BB to escape a cluster: (i) the center-of-mass (CoM) velocity of the new binary must exceed the local escape speed of the cluster and (ii) the binary must not undergo any strong (ionizing) encounters en route out of the cluster's gravity well. Unfortunately, the fraction of three-body encounters in which three unbound bodies encounter each other within a small volume, whilst simultaneously occupying the ``tail" of a cluster's velocity distribution, is vanishingly small. In fact, the outgoing binary CoM velocities from the 3UB simulations generated in \S~\ref{sec:The 3BBF Rate} are approximately described by a Maxwellian with a velocity dispersion ${\sim}70\%$ (smaller) than the median of the initial velocity dispersion. It is thus safe to assume that newly formed 3BBs do not immediately escape upon assembly. 

Instead of the ``immediate escape'' channel, we will consider the fraction of 3BBs which form in a cluster and survive within it until the time of the cluster's absolute dissolution. Our calculation employs the geometric cross-section, $\Sigma=\pi a^2$, rather than the ionization cross-section \citep{Hut_1983, HeggieHut1993} as we are only interested in binaries which do not undergo a single encounter over the rest of the cluster lifetime. We emphasize that this cross-section is strictly larger than the ionization cross-section (see equation 5.12 in \citet{Hut_1983}), and therefore serves as an upper bound.

The probability that a binary formed at time $t_{\rm form}$ survives unionized and without experiencing a single encounter until the cluster evaporates at $t=t_{\rm evap}$ is given by 
\begin{equation}\label{eq:enc rate}
    P_{\rm no \, enc} = \exp\left(-\int_{t_{\rm form}}^{t_{\rm evap}} \Gamma_{\rm enc}(t) \mathrm{d}t\right), \ \Gamma_{\rm enc} = 4 \sqrt{\pi} n \sigma a^2,
\end{equation}
integrated along the binary's orbit,  where $\Gamma_{\rm enc}$ is the local encounter rate. The rate depends on the radius of the binary's orbit in the cluster, $r(t)$, which itself varies with time due to the cluster's evaporation. This process occurs on much longer time-scales than the orbit of a binary in a cluster \citep{Gieles_2011}, so the orbit's actions are conserved. One of these actions is the angular momentum of the orbit, which, for a circular orbit is given by $l=2m \sqrt{GM_{\rm cl}({<}r) \, r}$. Thus, for such an orbit
\begin{equation}\label{eqn:adiabatic invariance orbit in cluster}
    r(t_{\rm form}) M_{\rm cl}({<}r(t_{\rm form});t_{\rm form}) = r(t) M_{\rm cl}({<}r(t);t),
\end{equation}
and the orbital radius expands to compensate for the lost mass of the cluster. As the radial action (which measures the radial fluctuations of the orbit; \citealt{Lynden-Bell1963,BT_2008}) is also conserved, the eccentricity of the binary's orbit about the cluster (defined as in \citealt{Lynden-Bell1963}) cannot change. For highly eccentric orbits, as well as for nearly circular ones, the periapsis distance $r_{\min}$ is proportional to the angular momentum. Therefore, we may take Equation~\eqref{eqn:adiabatic invariance orbit in cluster} to imply that the pericenter of the orbit grows with time, because the enclosed mass $M_{\rm cl}(<r)$ decreases; a converse process also occurs in the adiabatic contraction of halos \citep{BarnesWhite1984,Jesseitetal2002}.

\begin{figure*}
    \centering
    \includegraphics[width=\textwidth]{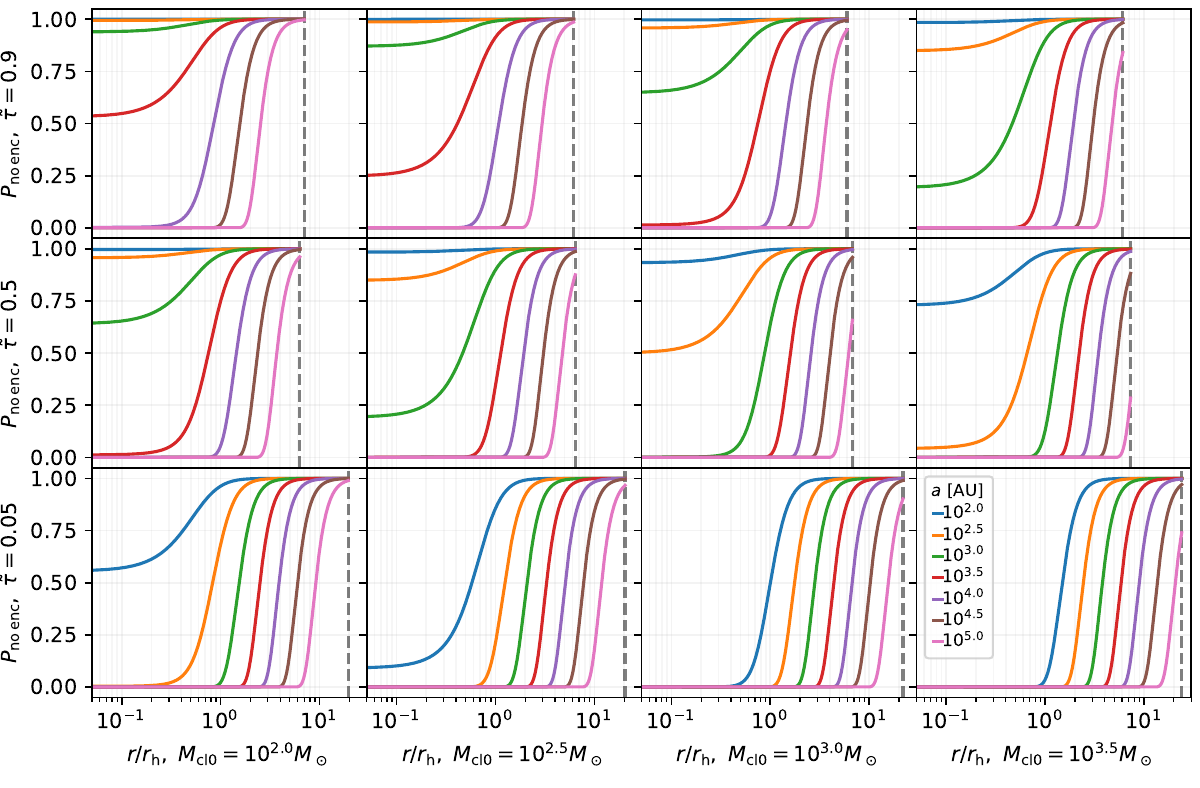}
    \caption{The probability a binary will experience no strong encounter for the rest of a clusters' life (Equation~\ref{eq:no enc}) versus the radial location in the cluster at the time the binary was born. The probability curves are organized such that columns correspond to initial cluster mass, $M_{\rm cl0}$, and rows correspond to the time since the cluster was born, $\tilde{\tau}$, in units of the total cluster lifetime. Binaries assembled with smaller SMA, later in a clusters life, or in lower mass clusters all have an elevated survival probability.}
    \label{fig:P_no_enc}
\end{figure*}

The encounter rate $\Gamma_{\rm enc}$ is proportional to the local number density, $n(r(t))$, and the local velocity dispersion, $\sigma(r(t))$, and thus depends on time via Equation~\eqref{eqn:adiabatic invariance orbit in cluster}. The quantity $n \, \sigma$ is a decreasing function of $r$, so $\Gamma_{\rm enc}(r_{\min})$ is the maximum value of the encounter rate along an orbit. As the orbit expands, this upper bound on $\Gamma_{\rm enc}$ decreases.  There are no explicit angle-action variables for a Plummer model, prohibiting the construction of an analytic formula for the time-dependent encounter rate $\Gamma_{\rm enc}(t)$. Instead, we seek a lower bound on the survival probability, $P_{\textrm{no enc}}$. Replacing $\Gamma_{\rm enc}(t)$ by its maximum initial value, $\Gamma_{\rm enc}(t_{\rm form})$, the survival probability becomes
\begin{equation}\label{eq:no enc}
\begin{aligned}
    P_{\rm no\, enc} &\geq \exp\left[- \Gamma_{\rm enc}(t_{\rm form}) \, (\tau_{\rm cl}-t_{\rm form})\right] \\&\approx \exp\left[- 4 \sqrt{\pi} \, n(r_{\rm form},t_{\rm form}) \, \sigma(r_{\rm form},t_{\rm form}) \, a^2 \, (\tau_{\rm cl}-t_{\rm form})\right].
\end{aligned}
\end{equation} 
In Figure~\ref{fig:P_no_enc}, $P_{\rm no \, enc}$ versus distance from cluster center, $r$, initial cluster masses ranging from $10^2$ to $10^{3.5} M_\odot$ and several key epochs in their evolution is displayed. For simplicity, we set the cluster orbital radius about the MW to be $R_G = 8 \, \rm{kpc}$, because most solar neighborhood 3BBs predicted by our model were emitted by clusters within $\pm 2 \rm kpc$ of the Sun's orbital radius. As expected, larger SMA have a smaller chance of surviving, while binaries assembled later in the clusters' life are less likely to be perturbed before dissolution.  

%Binary encounters between other wide binaries inhabiting an individual cluster are deemed negligible in our models because the binary-single encounter rate for wide binaries is always several orders magnitude larger than the wide-binary formation rate. This is explicitly illustrated by comparing Equations~\ref{eq:gammaf0} and \ref{eq:enc rate},
%\begin{equation}
%    \frac{\tilde{\Gamma}_{\rm F,0}}{\Gamma_{\rm enc}}\times  (4 \pi r^2 R_1) = \frac{12 \sqrt{2} c_0 \eta_H^2}{ N_{\rm cl}(t)^{2/3}} \frac{[r/b(t)]^2}{(1+ [r/b(t)]^2)^3}
%\end{equation}
%where the SMA of the 3BB is assumed of the order ${\sim}R_1$, which is a fraction of the local Hill radius, and the volume element is the spherical volume differential with the interaction volume radius as the radial element. If we choose $\eta_H=0.1$, there is no location or time in a cluster where the typical formation rate of a wide binary can compete with the destruction rate. This is not to say that encounters between two wide binaries cannot happen or influence the formation of exotic hierarchical systems, but these encounters are deemed statistically negligible for the purpose of our model.

Encounters between two wide binaries are deemed negligible in our models because the binary-single encounter rate for wide binaries is always much larger than the wide-binary formation rate.  This may be explicitly illustrated by, first, comparing the binary-binary and binary-single encounter rates,
\begin{equation}
	\frac{\Gamma_{\rm bb}}{\Gamma_{\rm enc}} \sim \frac{n_{\rm bin}\Sigma_{\rm bb}}{n \Sigma_{\rm enc}}\,.
\end{equation}
The cross-section for a strong binary--binary encounter is approximately $\Sigma_{\rm bb} \sim \pi (a_1 + a_2)^2$ \citep{Fregeau_2004}, while the cross-section for a strong binary-single encounter is $\Sigma_{\rm enc} \sim \pi a_1^2$. The steady-state number of wide binaries is suppressed relative to the number of single stars by a factor of the cluster size, i.e.~$n_{\rm bin}/n \sim 1/N(t_{\rm form})$, where $N(t_{\rm form})$ is the number of bodies in the host cluster at binary formation (see \citealt{BT_2008}, appendix M; \citealt{RoznerPerets2023}). Taking the SMA to be similar, we find
\begin{equation}
	\frac{\Gamma_{\rm bb}}{\Gamma_{\rm enc}} \sim  \frac{4}{N(t_{\rm form})} \ll 1\,.
\end{equation}
This relation approaches unity as the number of bodies in the cluster approaches $4$, at which point it is difficult to justify evolving the remaining system as a cluster or self-consistently calculating encounter rates. 

Encounters between wide and hard binaries, of which there can be many, may be treated identically to that with a single star because a wide binary's center of mass cannot significantly perturb a hard binary. Thus, we consider encounters between wide and hard binaries to be degenerate with binary-single encounters and included self-consistently within a Plummer cluster's density profile.

\section{Star Formation History and Radial Diffusion in the Milky Way's Disks}\label{sec:Milky Way, Broad}
Up until \S~\ref{sec:escape_disruption_cluster}, we only considered dynamics internal to toy Milky Way stellar clusters. To estimate the number and orbital element distributions of 3BB's in the solar neighborhood, it is critical to model the relationship between the clusters hosting 3BBF,  the MW disk, the cluster birth time, $t$, and Galactocentric radius, $R_G$, in the Milky Way thin and thick disks, and the journey (i.e., angular momentum diffusion) the few 3BB's managing to escape their home clusters must take to reach a comparatively small spatial region within ${\sim} 100\,\rm{pc}$ of the Sun. 

We may accomplish this by combining the thick disk model from \citet{Wagg_2022} with the updated thin disk model of \citet{Frankel_2020}; \citet{Wagg_2022} employed an older thin disk model. These models are a set of smooth semi-analytic distribution functions, tuned to reproduce the present day characteristics observed in the Milky Way. We highly recommend the reader carefully review their works for detailed information---only the set of distributions as implemented for our purposes are highlighted here.

The MW circular velocity curve is assumed to be constant across its total lifetime, $\tau_{\rm MW}= 12 \, \rm{Gyrs}$, and is taken exactly from the modeling scripts provided by \citet{Eilers_2019}. All references to the circular velocity as a function of the Galactocentric radius, $R_G$, should be interpreted as their full functional form, $v_c(R_G)$ (\textit{all stellar components + halo}), as displayed in Figure~1 of \citet{Eilers_2019}. Every orbiting body, whether stars or stellar clusters, are assumed to be on circular orbits. Escaping binary stars are the only objects we consider to be undergoing diffusion throughout the disk while clusters are evolved assuming fixed $R_G$. Finally, our toy galaxy is entirely composed of equal-mass, $0.5 M_\odot $ stars, as we are only interested in evaluating the feasibility of forming local field 3BBs, not an attempted reproduction of the exact stellar and compact object sub-populations observed in Gaia. 

\subsection{Milky Way Disk Evolution}\label{sec:Milky Way Evolution}
Restating our assumption that all stars are born in clusters, a star formation history model is, implicitly, a cluster formation history. The normalized set of equations describing the formation and dynamical evolution of bodies in the thick disk may be written
\begin{equation}
    \begin{aligned}
       &f_{ha}(R_G) = \frac{R_G}{R_{d0}^2} e^{-R_G/R_{d0}}, \\
        &SFH_{\rm ha}(t) = \begin{cases}
        \frac{e^{-t/\tau_{SFR}}}{\tau_{SFR} \,\left(1-e^{-\tau_{ha}/\tau_{SFR}}\right)}, & t<\tau_{ha}\\
          0, &t>\tau_{ha}
          \end{cases}\\
        &D_{\rm ha}(t', R_G, R_G') =  \sqrt{\frac{2 \tau_m}{\pi \, t'}} \exp\left[-\frac{\left(R_G' - R_G + \frac{c_1^2 \, t'/\tau_m}{2 R_{\rm d0} }\right)^2}{2c_1^2 \, t'/\tau_m} \right] {\rm Erfc}\left[ \frac{c_1^2 \, t'/\tau_m - 2 R_G R_{d0}}{2 \sqrt{2 \, t'/\tau_m}\, c_1 R_{\rm d0}} \right]^{-1},
    \end{aligned}
\end{equation}

\noindent where $R_{d0}=2.3 \,\rm kpc$ is the disk scale-length, $R_G$ is the Galactocentric distance of the cluster's orbit and the starting point from which escaping 3BBs begin their disk migration, $t$ is the time separation between the MW's birth (assumed to occur $\tau_{\rm MW}=12\, \rm Gyrs$ ago) and the cluster's birth, $\tau$ is the time between cluster birth and formation of the binary, $t'=\tau_{\rm MW}-\tau_{\rm cl}-t$ is the look-back time since the host cluster of a 3BB dissolves, and $R_G'$ is the location of the binary at $t'$. Additionally, $\tau_{SFR}=6.8 \, \rm Gyrs$ is the star formation timescale of the thick disk, $\tau_{ha}=6\,\rm Gyrs$ is the time when new cluster formation is ``turned off'' in the thick disk, $\tau_m=6 \, \rm Gyrs$ is a timescale controlling the diffusion strength in the disk, and $c_1=2.48 \, \rm{kpc}$ is the diffusion scale length. For simplicity, the latter two quantities and the form of the diffusion distribution, $D_{ha}$, are assumed identical to that of the thin disk models given by \citet{Frankel_2020}.

The equations describing the thin disk's evolution are mostly identical to the above equations---namely $f_{\rm la}$ and $D_{\rm la}$---with the caveat that $R_{d0}\rightarrow 2.8\, \rm kpc$ and that $SFH_{\rm la}(t)$ features ``inside out growth.'' In other words, the star formation rate is higher at earlier times closer to the MW center, while ``turning on'' at later times and farther out from the Galactic center. Ported from \citet{Frankel_2020} and normalized to a birth look-back time of $\tau_{la}=6 \,\rm Gyrs$, the star formation history of the thin disk is
\begin{equation}
\begin{aligned}
    SFH_{\rm la}(t, R_G) =&\begin{cases} \frac{y\,  e^{y (\tau_{\rm MW}-\tau_{la}-t)/\tau_{SFR}}}{\tau_{\rm SFR} \left(1- e^{-y \,\tau_{la}/\tau_{SFR}}\right)}, & \tau_{ha}<t<\tau_{MW}\\
    0, &\rm{otherwise} \end{cases}\\
    y = 1& -\frac{R_G}{12.6 \, \rm kpc},
\end{aligned}
\end{equation}
where $\tau_{la}=6 \,\rm\, Gyr$ and $\tau_{SFR}=1 \, \rm\, Gyr$. \citet{Wagg_2022} constructed their Milky Way model assuming a total stellar mass $M_d = 2.6 \times 10^{10} M_\odot$ in each of the thick and thin disks.  The number of stars in each present-day disk are $N_{\rm ha}=\frac{M_d}{m} = 7 \times 10^{10}$ and $N_{\rm la}=\frac{M_d }{m} = 5.2 \times 10^{10}$, respectively. By combining the two disk distributions, our evolving MW disk model is
\begin{equation}\label{eq:galactic evolution}
\begin{aligned}
    & \mathcal{DHL}(t, R_G, R_G') =  N_{\rm ha} f_{\rm ha}(R_G) SFH_{\rm ha}(t) D_{\rm ha}(t',R_G,R_G')+ N_{\rm la} f_{\rm la}(R_G) SFH_{\rm la}(t', R_G) D_{\rm la}(t',R_G,R_G').
\end{aligned}
\end{equation}

\subsection{The Cluster Initial Mass Function}\label{sec:cimf}
A cluster mass function (CMF) of the form $f(M_{\rm cl0})\propto M_{\rm cl0}^{-2} e^{-M_{\rm cl0}/M_c}$ is well-established observationally. The exponential (or \textit{Schechter}) cutoff applies a smooth means to regulate the maximum allowable cluster mass in a galaxy \citep{Schechter_1976, Elmegreen_2006, Gieles_2006,Zwart_2010, Just_2023}. \textit{When} in a galaxy's history and \textit{where} in a galactic environment a CMF is established remains a field of active inquiry, but progress has been made in connecting observed star formation rates to the cluster masses which preceded star formation epochs \citep{Gieles_2009,Johnson_2017,Choski_2021}.  Based on the preceding literature, we can construct a cluster initial mass function (CIMF) compatible with our MW evolution model (\S~\ref{sec:Milky Way Evolution}).

First, let's assert the form of the CIMF distribution as found by \citet{Just_2023}
\begin{equation}\label{eq:cimf}
    f_{\rm CIMF}(M_{\rm cl0}, t, R_G) \propto M_{\rm cl0}^{-1}(1+ M_{\rm cl0}^{2.4})^{-1/2} e^{-M_{\rm cl0}/M_c(t, R_G)}
\end{equation}
where $M_c(t,R_G)$ is the time- and Galactocentric radius-dependent cutoff of the Schechter function. \citet{Johnson_2017} empirically found that $M_c$ may be related to the star formation rate surface density, $\Sigma_{\rm SFR}$, with the expression

\begin{equation}\label{eq:Mc_sig_relation}
    \log_{10}\left(\frac{M_c}{M_\odot}\right) = 1.07 \log_{10}\left(\frac{\Sigma_{\rm SFR}}{M_\odot \rm{yr}^{-1} \rm{kpc}^{-2}}\right ).
\end{equation}
We choose to extend the dimensionality of the equation by additionally asserting that $\Sigma_{\rm SFR}\equiv\Sigma_{\rm SFR}(t,R_G)$. In our combined \citet{Frankel_2020, Wagg_2022} prescription, the Milky Way's $\Sigma_{\rm SFR}$ is implicitly

\begin{equation}
\begin{aligned}
    \Sigma_{\rm SFR}(t, R_G) &= \frac{M_d}{2 \pi R_G} \left(f_{\rm ha}(R_G) SFH_{\rm ha}(t) + f_{\rm la}(R_G) SFH_{\rm la}(t, R_G)\right),
\end{aligned}
\end{equation}
which is related to $M_c(t, R_G)$ using Equation~(\ref{eq:Mc_sig_relation}). The resulting normalization for Equation~(\ref{eq:cimf}) is numerically evaluated between $2 M_\odot$ \citep{Just_2023} and $M_c(t, R_G)$, since we consider $M_c$ to be the largest possible cluster mass in a star forming region \citep[see also][]{Choski_2021}.

\section{Field Evolution of Ejected Three-body Binaries}\label{sec:Field_evolution}
Binary evolution in the field is dependent on the internal stellar dynamics (e.g., tidal torquing \& binary stellar evolution) and the extrinsic MW secular dynamics and non-secular stellar encounters (e.g., orbital perturbations and disruptions). Our scope is limited to the case of typical (low-mass) stellar binaries, with separations large enough that the internal stellar dynamics are negligible on long time-scales, and only broadly care about the orbital properties of field 3BBs. Thus, our modeling of binary field evolution is solely dedicated to the extrinsic dynamical effects in the MW disk en route to the solar neighborhood.

Two works form the foundation of our field evolution prescriptions: \citet{Hamilton2022} and \citet{Hamilton_2024}. \citet{Hamilton2022} explored the MW secular dynamics and evolution of field binaries, a process named ``phase mixing,'' while \citet{Hamilton_2024} modeled the cumulative effect of many weak scattering encounters imparted to field binaries---``cumulative scatter'' for short. It is not straightforward to consider both of these field evolution prescriptions simultaneously, so we treat these works as bifurcating paths, both of which we explore. Application of \textit{phase mixing} (\textsc{PM}) is discussed in \S~\ref{sec:phasemix} and \textit{cumulative scatter} (\textsc{CS}) in \S~\ref{sec:cumulative scatter}.

\subsection{Binary Disruption and Phase Mixing}\label{sec:phasemixanddiskdisruption}

\subsubsection{Wide Binary Disruption in the Disk}\label{sec:disk disruption}
Binary disruption in the disk due to stellar encounters is not included \textit{a priori} through phase mixing. Appending a disruption survival probability, $P_{\rm d, MW}$ (Equation~\ref{eq:disk disruption probability}), to our final equation is an easy fix. The final equation, Equation~\ref{eq:G3R}, returns the total number of binaries when integrated, but with initially unevolved binary samples. Their orbital elements remain identical to what was assigned at formation in the cluster---it is to this final distribution of surviving binaries we apply phase mixing, as described in \S~\ref{sec:phasemix}.\footnote{Note that phase mixing only modifies the eccentricity distribution of a binary population, leaving the SMA distribution unmodified.}

The fraction of binaries which ``diffuse'' to a present-day Galactocentric orbit, $R_G'$, will naturally be curtailed by strong scattering encounters in the disk. It is naturally preferable to consider an evolving disk model, accounting for the changing number and density of bodies in the disk with look-back time, $t'$ (i.e., the time the binary leaves it's dissolving natal cluster). To avoid numerous intractable difficulties in the calculation, we instead adopt the conservative simplifying  assumption of a constant disk density equal to the present-day density distribution.  

A disk-survival model may be constructed by first writing the classic disruption rate \citep{BT_2008} 
\begin{equation}
\begin{aligned}
    &\Gamma_d(R_G) = \frac{22.7\, G \,a \ln (\Lambda_d)}{\sigma_{\rm MW}} \rho_{\rm MW}(R_G), \ \Lambda_d= \frac{a \sigma_{\rm MW}}{3 G m},
\end{aligned}
\end{equation}
where $\rho_{\rm MW}(R_G)$ is the mass density of stars at $R_G$, $a$ is the SMA of the binary, $\sigma_{\rm MW}$ is the relative velocity dispersion between bodies in the disk, and $\ln(\Lambda_d)$ is the Coulomb logarithm within the disk with $m = 0.5 M_\odot$. For simplicity, we assume a constant $\sigma_{\rm MW}=41 \, \rm{km/s}$ throughout the disk. The probability of a binary not being disrupted in the disk is then
\begin{equation}\label{eq:disk disruption probability}
\begin{aligned}
    &P_{\rm d,MW}(a, t, \tau, R_G, R_G')= e^{-\lambda_{\rm d,MW}},\\
    &\lambda_{\rm d,MW} = \frac{1}{v_d}\int^{R_G'}_{R_G} \Gamma_d(R_G) dR_G = \frac{22.7 \,G a \ln(\Lambda_d)}{\sigma_{\rm MW} v_d} \int^{R_G'}_{R_G}\rho_{MW}(R_G) dR_G,\\
    &v_d=\frac{R_G'-R_G}{t'}, \ t'= \tau_{\rm MW} - \tau_{\rm cl}- t
\end{aligned}
\end{equation}
where $v_d$ is the average ``diffusion velocity'' and $\tau_{\rm cl}$ is the total lifetime of the cluster.

Following \citet{Eilers_2019}, we model the present-day thick and thin disk mass densities with the \citet{Miyamoto_1975} profile and the bulge with a simple Plummer density profile. Since vertical travel in the disk is not considered, only in-plane motion, we treat the binary as if it only travels within the central (and densest) portions of Milky Way. With these simplifications, the integral of Equation~(\ref{eq:disk disruption probability}) becomes
\begin{equation}\label{eq:disk_density}
    \begin{aligned}
        \int^{R_G'}_{R_G}\rho_{\rm MW}(R_G) dR_G = & \ \sigma_{\rm disk}(M_d,R_G', c_{\rm ha},b_{\rm ha})- \sigma_{\rm disk}(M_d,R_G, c_{\rm ha},b_{\rm ha})\\ 
        &+ \sigma_{\rm disk}(M_d,R_G',c_{\rm la},b_{\rm la})- \sigma_{\rm disk}(M_d,R_G,c_{\rm la},b_{\rm la})\\
        &+ \sigma_{\rm buldge}(M_d, R_G', b_b) - \sigma_{\rm buldge}(M_d, R_G, b_b),\\
        \sigma_{\rm disk}(M,R,c,b) =& \frac{M R}{4 \pi c^3} \frac{1+(1+c/b) \left[1+(R/c)^2\right]}{\left[1+(R/c)^2\right]^{3/2}}, \ \sigma_{\rm buldge}(M, R, b) = \frac{3 M R}{4 \pi b^3} \frac{1 + \frac{2}{3}\left(R/b\right)^2}{\left[1 + (R/b)^2 \right]^{3/2}},
    \end{aligned}
\end{equation}
with the disk mass $M_d=2.6\times10^{10}M_\odot$ for both disks, thin disk distance scales $(c_{\rm la}, b_{\rm la}) = (5.55 \, \rm{kpc}, 0.25 \, \rm{kpc})$, thick disk distance scales $(c_{\rm ha}, b_{\rm ha}) = (3.4 \, \rm{kpc}, 0.8 \, \rm{kpc})$, bulge mass $M_b=0.9 \times 10^{10} M_\odot$, and bulge Plummer scale $b_b=0.3 \, \rm{kpc}$.

\subsubsection{Phase Mixing}
\label{sec:phasemix}
After a binary escapes a cluster, if the binary is sufficiently small not to be destroyed by Galactic tides (i.e.~if $a \lesssim 2\times 10^5$ AU), the binary evolves secularly under their influence \citep{HeislerTremaine1986,HamiltonRafokiv2019I,HamiltonRafokiv2019II,Hamilton2022,GrishinPerets2022}. \cite{Hamilton2022} showed that the consequence of such evolution is a ``phase mixing'' (\textsc{PM}) in the appropriate phase-space region. This secular evolution is described by the Galactic-tide Hamiltonian $\ham$, given by \citep{HamiltonRafokiv2019I,HamiltonRafokiv2019II}
\begin{equation}
\begin{aligned}
    \ham & = \frac{\kappa^2 a^2 \mu}{8}\Big[\left(2 + 3 \epsilon^2\right)\left(1-3\Gamma \cos^2i\right) -15\Gamma \epsilon^2 \sin^2 i \cos 2\omega\Big]
\end{aligned}
\end{equation}
in the test-particle, quadrupole approximation, where $\kappa$ is a combination of the epicycle frequencies of the orbit of the binary in the Galaxy, and $\Gamma$ is a ratio encoding the tidal perturbations from an axisymmetric tidal tensor; we use $\Gamma = 1/3$ for concreteness, but our results are insensitive to the precise value. Observe, that $\ham$ is orbit-averaged, and conserves the (normalized) $\zhat$ component of the angular momentum $j_z \equiv \cos i\sqrt{1-\epsilon^2}$, for the argument of the ascending node is a cyclic co-ordinate. Thus, $\ham$ describes a two-dimensional phase-space, parameterized by the eccentricity $\epsilon$ and the argument of pericenter $\omega$, and admits two constants of motion: $\ham$ itself, and $j_z$. 

If the initial distribution is $f(a,\epsilon)$, then \textsc{PM} implies that $f$ is ``smeared out'' along the contours of constant $(\ham,j_z)$. So, in the long-time limit, the final distribution is just $f$, averaged appropriately over these contours, as shown by \cite{Hamilton2022,ModakHamilton2023}; we sketch the procedure here for completeness. Assuming that the initial distribution $f$ is independent of $i$, $\omega$ and $\Omega$---because triple encounters are isotropic---we express $f$ in terms of the appropriate action variables, $j \equiv \sqrt{1-\epsilon^2}$ and $j_z$, \emph{viz.}
\begin{equation}
    f_{\rm A}(j,j_z) \equiv \frac{j}{\sqrt{1-j^2}} f\left(a,\sqrt{1-j^2}\right),
\end{equation}
(where we have suppressed the $a$-dependence on the left) and then calculate the resultant ``energy'' distribution, $f_{E}(\ham,j_z)$, as explained in equation (A1) of \cite{Hamilton2022} (which is denoted $f_\infty(\mathbf{w})$ there). The final distribution is then 
\begin{equation}
    f_\infty(j) \propto \int_0^{2\pi} \mathrm{d}\omega \int_{-j}^j \mathrm{d}j_z ~ f_{E}(\ham(j,j_z,\omega),j_z),
\end{equation}
where the proportionality coefficient is fixed by normalizing $\int_0^1 f_\infty(j)\mathrm{d}j = 1$. Converting back to eccentricity yields that the phase-mixed distribution of binaries is $f'(a,\epsilon)= \epsilon f_\infty\left(\sqrt{1-\epsilon^2}\right)/\sqrt{1-\epsilon^2}$.

\subsection{Cumulative Scatter}\label{sec:cumulative scatter}
\citet{Hamilton_2024} present a powerful framework which simultaneously evolves the SMA and eccentricity of field binaries through the cumulative effect of many scattering encounters---here shortened to ``cumulative scatter'' (\textsc{CS})---while tracking their eventual disruption, should one occur. Unlike with the \textsc{PM} mode of field evolution (\S~\ref{sec:phasemixanddiskdisruption}, we do not need to include the probability of binary disruption en route from the natal cluster (\S~\ref{sec:escape_disruption_cluster}) to the solar neighborhood as it diffuses through the disk (\S~\ref{sec:Milky Way Evolution}). Instead, samples drawn by \textsc{emcee} are evolved in post-processing, removing samples ending in disruption through \textsc{CS}. 

Post-escape field evolution is governed by the Fokker-Planck equation, with the binary samples evolved using the same Monte-Carlo method described in \S~3.4 of \citet{Hamilton_2024}. For the readers convenience, we rewrite their Euler-Maruyama update rules here:
\begin{equation}\label{eq:fokker_planck_update}
    \begin{aligned}
        a'&= a + \delta a, \  \epsilon'^2 = \epsilon^2 + \delta (\epsilon^2), \ t' = t + \delta t, \, R_G' = R_G + v_d \delta t\\
        \delta a &= A_a \delta t + \xi_a  (D_a \delta t)^{1/2}, \ \delta (\epsilon^2) = A_{\epsilon} \delta t + \xi_\epsilon  (D_{\epsilon} \delta t)^{1/2},\\
        A_a &= \frac{7 \langle (\Delta v)^2\rangle}{3 G m_b}a^2, D_a =\frac{4 a}{7} A_a,\  A_e = \frac{5}{7} A_a (1 - 2 \epsilon^2), \ D_\epsilon = \frac{10}{7} A_a \epsilon^2 (1-\epsilon^2), \\
         \langle (\Delta v)^2\rangle &= \frac{16 \sqrt{2 \pi}G^2 m_p}{\sigma_{MW}} \rho_{MW}\left(R_G'\right) \,  \ln\left(\frac{a \, \sigma_{\rm MW}^2}{3 G m_p}\right),
    \end{aligned}
\end{equation}
where $(a', \epsilon'^2)$ are the updated binary properties during time step $\delta t$, $m_b=1 M_\odot$ is the total mass of the binary, $m_p=0.5 M_\odot$ is the average stellar mass, $v_d$ is the average radial diffusion velocity of the binary---see also Equation~(\ref{eq:disk disruption probability})---and $(\xi_a,\xi_\epsilon)$ are independent Gaussian random numbers with mean 0 and variance 1.\footnote{Note that in \citet{Hamilton_2024}, $\epsilon$ is the square of the binary eccentricity, while this work employs $\epsilon$ as the eccentricity due to the symbolic degeneracy with the Euler's number.} 

The term $\rho_{MW}(R_G)$ is the present-day stellar density of the MW at Galactocentric radius, $R_G$; the functional definition may be found by taking the first derivative in $R$ of Equation~(\ref{eq:disk_density}). Just as in \S~\ref{sec:disk disruption}, we hold the disk stellar density profile constant to simplify the calculation of disk disruption and only evolve the binary assuming a circular orbit in a smooth and continuous radial diffusion through the disk to the solar neighborhood. 

Numerical accuracy and efficiency is optimized by employing time-regularization, calculating each time-step to be $1\%$ the minimum of the ``naive timescales'' described in \citet{Hamilton_2024}. Expressed mathematically, 
\begin{equation}\label{eq:deltat Hamilton}
    \delta t = 0.01 \times \min\left(\frac{a}{A_a}, \frac{\epsilon^2}{A_\epsilon}\right).
\end{equation}

Finally, we include a fixed probability of binary disruption due to diffusive interactions with giant molecular clouds (GMC) in every time-step. A simple approximation for the GMC disruption rate may be found in \citet{BT_2008},
\begin{equation}
    \Gamma_{\rm GMC} = \frac{(a/10^4 \rm{AU})^3}{190 \rm{Gyr}}, 
\end{equation}
where the expression has been simplified to the case of $1 M_\odot$ binaries with a relative velocity dispersion between GMCs and disk stars of $\sim 30 \rm{km\,s^{-1}}$. The probability that a binary evolving under \textsc{CS} survives interactions with GMCs in each time-step is
\begin{equation}\label{eq:Pdgmc}
    P_{\rm d,GMC} = e^{-\Gamma_{\rm GMC} \delta t}
\end{equation}
where $\delta t$ is the regularized time-step determined in Equation~(\ref{eq:deltat Hamilton}). Should a binary's SMA increase beyond $1 \rm pc$ or a random, uniformly drawn number between 0 and 1 is greater than Equation~(\ref{eq:Pdgmc}), the binary is considered destroyed and removed from the \textsc{emcee} samples.

\begin{table*}
\centering
\caption{\textbf{G3R Integration Bounds}}
\label{tab:integration bounds}
  \centering
  \begin{tabular}{lccl}
    \hline \hline
    Symbol & Min & Max & Description \\
    \hline
    $\tilde{r}$      &  0               & 1                         & the center of a cluster to its tidal boundary\\
    $\tilde{\tau}$   &  0               & 1                         &  the beginning of cluster evolution to dissolution\\
    $M_{\rm cl0}$    &  $50 \ M_\odot$& $10^6 \ M_\odot    $& an initial mass containing 100 bodies to a mass encompassing ${\gg}99\%$ of the parameter space\\
    $R_G$            &  $0.5 \ \rm{kpc}$  &  $20 \ \rm{kpc}           $ &  a region near the center of the MW to an arbitrarily large radius\\
    $R_G'$           &  $7.9 \ \rm kpc $& $8.1\ \rm kpc                $& the subset of final orbital radii within 100 pc of the sun\\
    $t$              &  $0 \ \rm{Gyrs} $  & $12 \ \rm{Gyrs}            $& the entire lifetime of the MW\\
    $\epsilon$       &  0               & 1                         & all possible eccentricities\\
    $a$              &  $1 \ \rm{AU}$     & $1 \ \rm{pc}              $ & tight 1 AU binaries to binaries as wide as the largest detected in Gaia ($\sim 1$ pc)\\
    $k$              &  0               & 3                         & up to a final value of $k$ arbitrarily large enough to encompass $\gg99\%$ of the parameter space \\
    \hline
  \end{tabular}
\end{table*}

\section{The Galactic 3BBF Rate}\label{sec:G3R}
Having assembled the creation, diffusion, and destruction portions of the final distribution, we are finally in a position to discuss subtleties of the total integration. The final equation, our ``Galactic 3BBF Rate,'' (G3R) is the product of Equations~\ref{eq:gamma_plum_ex}, \ref{eq:UEE}, \ref{eq:no enc},  \ref{eq:galactic evolution}, \ref{eq:cimf}, and \ref{eq:disk disruption probability}, built with the numerous intermittent expressions. Written in functional form, the G3R is
\begin{equation}\label{eq:G3R}
    \begin{aligned}
        &\mathcal{G}_{\rm MW}d\Omega = \frac{\volume_*}{\volume_{\rm R_{G,s}}} f_{\rm CIMF}(M_{\rm cl0}, t, R_G)\mathcal{C}(M_{\rm cl0}, \tilde{\tau}, R_G, r, k, \epsilon, a) P_{\rm no \, enc}(M_{\rm cl0}, \tilde{\tau}, R_G, r,a) \\
        &\qquad \qquad \times\mathcal{DHL}(t, \tau, R_G, R_G') P_{\rm d, MW}(a, t, \tau, R_G, R_G')d\Omega,\\
        &d\Omega= r_{\rm t}\left(M_{\rm cl0}, \tilde{\tau}, R_G \right) \, \tau_{\rm cl}(M_{\rm cl0}) d\tilde{r} \, d\tilde{\tau} \, dk \, d\epsilon \, da \, dM_{\rm cl0} \, dR_G  \, dR_G' \, dt \, 
    \end{aligned}
\end{equation}
where $\volume_*/\volume_{\rm R_{G,s}} =\delta l^2/(3 R_{\rm G,s} h)= 4.2\times 10^{-4}$ is the fraction of a $1 \rm kpc$ thick disk annulus at $R_{\rm G,s}=8 \rm kpc$ within a $\pm 100$ pc radius of the Sun, and the substitutions $r = r_t \, \tilde{r}$, $dr= r_t d\tilde{r}$, $\tau = \tau_{\rm cl} \, \tilde{\tau}$, and $d\tau = \tau_{\rm cl} d\tilde{\tau}$ are made for a more convenient integration experience. 

The above expression includes $P_{\rm d, MW}$, but is only applicable when evaluating the G3R with \textsc{PM} (\S~\ref{sec:phasemixanddiskdisruption}); when using the \textsc{CS} prescription (\S~\ref{sec:cumulative scatter}), it is removed. Instead, the fraction of \textsc{emcee} samples eliminated in \textsc{CS} post-processing serves as a reduction factor applied to the G3R. 

The integration bounds are detailed in Table~\ref{tab:integration bounds}. A number of additional constraints are applied to Equation~(\ref{eq:G3R}), setting it to $0$ when triggered. These constraints include:
\begin{enumerate}
    \item $N_{\rm cl}(\tilde{\tau})\leq 20$; the validity of a Plummer model is contingent to many bodies being present. We arbitrarily select $N=20$ as a break point for integration.  
    \item $a\left(1-\epsilon\right)<0.1 \rm{AU}$; physical collisions and strong decay by tidal forces before and after 3BBF would prevent tight passages.
    \item $\tau_{\rm MW}-\tau_{\rm cl}-t<0$; this only occurs if the migration timescale is larger than the time a binary has to migrate to $R_G'$ from $R_G$ by the present day.
\end{enumerate}

Incorporating all stated prior conditions with Equation~(\ref{eq:G3R}), the G3R is integrated over the entire parameter space using a custom written Monte Carlo integration scheme which divides the parameter space into $\sim 100$ sub-volumes and draws a total of $10^9$ samples, resolving the integral to high-precision. We additionally employ \textsc{emcee} \citep{emcee} to uncover the underlying nine-dimensional distribution function accompanied by the same restrictions quoted above.

% \clearpage

\bibliography{binaries,encounters}{}
\bibliographystyle{aasjournal}



\end{document}